\documentclass[preprint,11pt]{elsarticle}
\usepackage{etoolbox}
\usepackage[margin=1in]{geometry}
\apptocmd{\thebibliography}{\raggedright}{}{}
\usepackage{comment}
\usepackage{endnotes}
\usepackage{lscape}
\usepackage{subcaption}
\usepackage{threeparttable}

\usepackage{array}
\newcolumntype{C}[1]{>{\centering\let\newline\\\arraybackslash\hspace{0pt}}m{#1}}
\newcolumntype{L}[1]{>{\raggedleft\let\newline\\\arraybackslash\hspace{0pt}}m{#1}}
\newcolumntype{R}[1]{>{\raggedright\let\newline\\\arraybackslash\hspace{0pt}}m{#1}}
 \usepackage{graphicx}
 \usepackage{subcaption}
 \usepackage{adjustbox}
\usepackage{algorithm}
\usepackage{algpseudocode}
 \usepackage{multirow}
 \usepackage{multicol}
\usepackage{atbegshi}% http://ctan.org/pkg/atbegshi
\AtBeginDocument{\AtBeginShipoutNext{\AtBeginShipoutDiscard}}
\biboptions{authoryear}

\usepackage{amssymb}
\usepackage{amsmath}
\usepackage{amsthm}
\usepackage{color}
\usepackage[normalem]{ulem}
\usepackage{hyperref}
\usepackage{setspace}
\theoremstyle{plain}% Theorem-like structures provided by amsthm.sty
\usepackage{amsthm}

\theoremstyle{definition}

\theoremstyle{remark}

\journal{}

\begin{document}
	\emergencystretch 3.5em

\begin{frontmatter}

\title{Impact of Transmission Dynamics and Treatment Uptake, Frequency and Timing on the Cost-effectiveness of Directly Acting Antivirals for Hepatitis C Virus Infection}
\label{editor_v3}

\author[inst1]{Soham Das}

\author[inst2]{Ajit Sood}
\author[inst3]{Vandana Midha}
\author[inst2]{Arshdeep Singh}
\author[inst4]{Pranjl Sharma}
\author[inst1]{Varun Ramamohan\corref{cor1}}

\cormark[cor1]
\ead{varunr@mech.iitd.ac.in}
\cortext[mycorrespondingauthor]{Corresponding Author}

\affiliation[inst1]{organization={Department of Mechanical Engineering, Indian Institute of Technology Delhi},%Department and Organization
	addressline={Hauz Khas}, 
	city={New Delhi},
	postcode={110016}, 
	state={Delhi},
	country={India}}

\affiliation[inst2]{organization={Department of Gastroenterology},%Department and Organization
            addressline={Dayanand Medical College and Hospital}, 
            city={Ludhiana},
            postcode={141001}, 
            state={Punjab},
            country={India}}
            
            \affiliation[inst3]{organization={Department of Medicine},%Department and Organization
            	addressline={Dayanand Medical College and Hospital}, 
            	city={Ludhiana},
            	postcode={141001}, 
            	state={Punjab},
            	country={India}}
 
            	\affiliation[inst4]{organization={Department of Community Medicine},%Department and Organization
            		addressline={Dayanand Medical College and Hospital}, 
            		city={Ludhiana},
            		postcode={141001}, 
            		state={Punjab},
            		country={India}}
\setstretch{1.5}
\begin{abstract}
%% Text of abstract
Cost-effectiveness analyses, based on decision-analytic models of disease progression and treatment, are routinely used to assess the economic value of a new intervention and consequently inform reimbursement decisions for the intervention. Many decision-analytic models developed to assess the economic value of highly effective directly acting antiviral (DAA) treatments for the hepatitis C virus (HCV) infection do not incorporate the transmission dynamics of HCV, accounting for which is required to estimate the number of downstream infections prevented by curing an infection. In this study, we develop and validate a comprehensive agent-based simulation (ABS) model of HCV transmission dynamics in the Indian context and use it to: (a) quantify the extent to which the cost-effectiveness of a DAA is underestimated - as a function of its uptake rate - if disease transmission dynamics are not considered in a cost-effectiveness analysis model; and (b) quantify the impact of the frequency and timing of treatment within a disease surveillance period, also as a function of uptake rate, on DAA cost-effectiveness. %The process of accomplishing the above research objectives also motivated the development of a novel random sampling and allocation approach, along with associated theoretical grounding, to estimate individual-level outcomes within an ABS that incurs substantially lower computational expense than the benchmark incremental accumulation approach. 

\end{abstract}

%%Graphical abstract
% \begin{graphicalabstract}
% \includegraphics{grabs}
% \end{graphicalabstract}

%%Research highlights
% \begin{highlights}
% \item Research highlight 1
% \item Research highlight 2
% \end{highlights}

\begin{keyword}
%% keywords here, in the form: keyword \sep keyword
OR in developing countries, Hepatitis C virus, Agent-based simulation, Transmission dynamics, Treatment frequency and timing
%% PACS codes here, in the form: \PACS code \sep code
%\PACS 0000 \sep 1111
%% MSC codes here, in the form: \MSC code \sep code
%% or \MSC[2008] code \sep code (2000 is the default)
%\MSC 0000 \sep 1111
\end{keyword}

\end{frontmatter}

%% \linenumbers
\setstretch{1.5}
%% main text
\section{Introduction}
\label{intro}

The hepatitis C virus (HCV) infection is the leading cause of liver cirrhosis and hepatocellular carcinoma (HCC), and globally affects as many as 58 million people \citep{who24}. The global prevalence of HCV is approximately 1\% \citep{cdc24}; however, it is much higher in many developing countries. For example, it is 11.6\% in Pakistan, 5.9\% in Gabon in Africa, and 3.0\% in Uzbekistan \citep{cdc24}. In India, where our study is situated, the prevalence of HCV was estimated at 0.5\%-1.5\% \citep{sood2018burden}. However, there is considerable variation in the prevalence of HCV within India, with certain states (provinces) such as Punjab reporting a prevalence as high as 3.6\% \citep{sood2018burden}. In such regions, poor access to screening and treatment has been identified as a key contributor to mortality from the disease \citep{who24}. Thus a comprehensive effort to improve HCV outcomes in these regions would involve accurately characterizing the cost-effectiveness of HCV treatment, and then quantifying the health and economic impact of the timing and frequency of HCV screening and treatment programs so that their cost-effectiveness can be optimized. In this study, we accomplish this via an agent-based simulation of HCV transmission dynamics situated in the Indian state of Punjab.

Cost-effectiveness analyses are routinely used across the world for comparing interventions for health conditions, and in particular, for informing reimbursement decisions for new interventions (such as a new drug or a new disease screening policy) made by payers such as public health agencies (e.g., the United Kingdom) or insurance companies (e.g., in the United States). Such analyses quantify the health and cost value of an intervention - typically in terms of quality-adjusted life years (QALYs) and costs - via the use of decision-analytic models such as Markov chains, decision trees, discrete-event or agent-based simulations \citep{chhatwal2016systematic}. In the context of HCV, such analyses have been used to, for example, determine optimal policies for prioritizing HCV treatment in US prisons \citep{ayer2019prioritizing}, and to quantify and summarize the health and cost benefits of universal screening and treatment with next-generation directly acting antiviral (DAA) drug regimens in India \citep{chugh2021cost}. Models employed in cost-effectiveness analyses of interventions for infectious diseases ideally must consider transmissions of infections, so that the population-level health and cost benefits of \textit{preventing} new infections, in addition to those of \textit{curing} existing infections, are also considered when making the reimbursement decision. However, as documented in literature reviews of cost-effectiveness analyses conducted for HCV \citep{chhatwal2016systematic} and HIV/AIDS \citep{gopalappa2017combinations}, very few such models incorporate disease transmission. This is the case for cost-effectiveness analyses conducted in the Indian context as well. In fact, in our knowledge, a comprehensive model of HCV transmission dynamics that considers the inherent stochasticity of infectious disease transmission has not yet been developed for the Indian - and indeed for the developing world - contexts.%Not considering transmissions in cost-effectiveness analyses can substantially underestimate the health and cost benefits of interventions such as treatment \citep{chhatwal2016systematic}. %%comparing public health interventions (such as opioid substitution programs) targeted at injecting drug users (IDUs) in Eastern Europe and Central Asia \citep{mabileau2018intervention}

The extent to which HCV treatment prevents new infections, and indeed is cost-effective, depends on its uptake rate - that is, the proportion of those infected who complete the treatment regimen. However, the marginal costs of expanding treatment to more infected patients may increase substantially beyond a certain threshold uptake rate. Therefore, given the above context, the following research questions arise: (a) quantitatively, what is the impact of \textit{not} incorporating transmissions on the cost-effectiveness of treatment; (b) what is the impact of uptake rate on the cost-effectiveness of treatment, and on the underestimation of health and cost benefits when transmissions are not considered; and (c) how can we determine the threshold uptake rate beyond which expanding reimbursable access to treatment is no longer cost-effective? We address these research questions as part of our first research objective ($RO1$): \textit{quantify the impact of incorporating HCV transmission dynamics on the epidemiological, health and cost outcomes associated with HCV treatment as a function of its uptake rate}.

HCV is a disease with slow progression, with patients typically spending long durations in the initial fibrotic stages before progressing to the cirrhotic stages or developing hepatocellular carcinoma. This implies that under resource-constrained conditions that characterize many developing countries, the timing and frequency of screening and treatment programmes for HCV may not necessarily need to be similar to that of other diseases (even HIV). For example, annual frequencies may not be cost-effective in terms of both health and economic outcomes. Further, the timing of deployment of interventions may also need careful consideration. For instance, consider a disease surveillance programme initiated at $T_0$ and slated to run until $T_N$. If resource constraints allow organizing one screening and treatment camp during the surveillance programme, then the question arises as to whether running the camp at $T_0$ would be more beneficial in terms of health and cost outcomes than at $T_N$, given the slow progression and (relatively) slow spread of the disease. Similarly, resources permitting, would running three camps be more cost-effective than two? If not, for the two-camp policy, when should they be held during the surveillance period? We address these questions as part of our second research objective ($RO2$): \textit{quantitatively characterize the impact of timing and frequency of HCV treatment, as a function of its uptake rate, with respect to a disease surveillance period.}  

In our knowledge, these questions have not previously been addressed in the extant literature. The answers to these questions may find application beyond just HCV: for example, they have not in our knowledge also been addressed for HIV or TB via a comprehensive modeling study such as ours - in particular, not in the developing world. %Further, our development of a comprehensive model of HCV transmission dynamics in the developing world also represents a key research contribution. This is because, unlike in the developed world where the dominant mode of HCV transmission is via injecting drug use, unsafe medical procedures represent the dominant mode of HCV transmission in many developing countries \citep{chakravarti2013study,farhan2023national}. Thus our ABM is the first to explicitly model disease transmission modes in addition to injecting drug use.

The remainder of the paper is organized as follows. In Section \ref{sec:litreview}, we discuss the relevant literature and describe our research contributions. In Section \ref{sec3}, we describe ABM development. In Section \ref{numres}, we discuss computational experiments conducted to address $RO1$ and $RO2$, and we conclude in Section \ref{sec:disc} with a discussion of the potential impact of our study and avenues for future research.

\section{Literature Review}
\label{sec:litreview}
In this section, we discuss the literature relevant to research objectives \textit{RO1} and \textit{RO2}, and discuss our research contributions with respect to each objective. Prior to this, we provide a brief overview of the different types of models that have been developed for assessing the epidemiological, health and economic impact of HCV treatment and/or screening, with a focus on treatment with DAAs. Further, in Table~\ref{tabstudies} below, we describe the type of transmission model, transmission modes considered, type of disease progression model, the geography of analysis, whether single or multiple treatment uptake rates and frequencies were considered, whether timing of treatment was considered, the outcomes estimation approach, and the types of outcomes generated by the model. In other words, this table summarizes the literature relevant to all three research objectives.

HCV treatment and/or screening impact assessment models are either static or dynamic models. Static models assume that the force of infection associated with HCV is constant throughout the period of evaluation, whereas dynamic models assume that the force of infection at each point in time depends upon the number infected at that time point \citep{brennan2006taxonomy,kim2008cost}. Static models include state-transition models (e.g., Markov models) or individual-level microsimulations. Dynamic models include (a) aggregate-level compartmental models wherein transmission dynamics are captured via differential equations, and (b) individual-level simulations, which can include Monte Carlo microsimulations, discrete-event simulations, and ABMs. A systematic review of model-based cost-effectiveness analyses for HCV published in 2016 \citep{chhatwal2016systematic} revealed that until then, only static models had been developed for this purpose. However, a number of dynamic models, both compartmental and agent-based, have been developed since the publication of this review. 

Given that HCV is an infectious disease, a dynamic model may best capture the impact of an anti-HCV intervention. A number of compartmental differential equation based models have been developed for this purpose, as will be discussed in the next subsection. However, such models do not capture the impact of individual-level heterogeneity on disease-related outcomes, whether at the population or patient level. Further, estimating the uncertainty in model outcomes by incorporating stochasticity into such models is a non-trivial task. These drawbacks are addressed by ABMs, wherein disease spread is modelled via interactions between entities (referred to as agents) that represent members of the population of interest \citep{willem2015agent}. ABMs provide a flexible and comprehensive approach for modelling the dynamics of infectious disease transmission at a sufficiently detailed level. A key challenge associated with the development of ABMs includes availability of reliable data for parameterizing agent interaction models in the environments of interest \citep{abualkhair2020managing,ghaderi2022public}.

A limited number of agent-based models have been developed for cost-effectiveness analyses for anti-HCV interventions. Extant ABMs have been developed for the United States \citep{he2016prevention,ayer2019prioritizing}, France \citep{cousien2015dynamic} and the Netherlands \citep{van2016cost}. These models focus on HCV transmission among injecting drug users (IDUs). However, in the Indian, and indeed in the broader South Asian and developing world contexts, the bulk of HCV transmissions occur via contaminated medical procedures such as blood transfusions \citep{lim2020effects,das2019abm,farhan2023national}. Building upon preliminary work \citep{das2019abm}, this study presents the first ABM developed for the developing world, and indeed the first that explicitly models HCV transmission through modes other than injecting drug use. 

Table~\ref{tabstudies} summarizes the dynamic transmission studies developed for HCV. 

%\begin{landscape}
\begin{table}[htbp]
	\centering
	\caption{Summary of dynamic transmission models developed to assess the impact of DAA treatment strategies for HCV. \textit{Abbreviations and acronyms.} S: Single, M: Multiple; W: Within horizon, L: Lifetime horizon, F: Fixed horizon; M1: Injecting drug use, M2: sexual route, M3: Implicit non-injecting drug use, M4: Implicit modelling through risk values, M5: Medical procedures, M6: Community risk, M7: Hemodialysis.}
	\resizebox{\textwidth}{!}{
		\begin{tabular}{|C{4cm}|c|C{2cm}|c|C{2cm}|C{2cm}|c|c|C{2cm}|C{2cm}|}
			\hline
			&       &       &       &       &       & \textbf{Screening/} & \textbf{Timing of} &       & \textbf{Types of} \\
			&       &       & \textbf{Disease} &       &       & \textbf{treatment} & \textbf{treatment} & \textbf{Horizon of health and} & \textbf{outcomes} \\
			& \textbf{Transmission} & \textbf{Transmission} & \textbf{progression} & \textbf{Geography} & \textbf{Uptake rate(s)} & \textbf{frequencies} & \textbf{considered} & \textbf{cost outcomes} & \textbf{evaluated} \\
			\textbf{Study} & \textbf{model} & \textbf{modes} & \textbf{model} & \textbf{of analysis} & \textbf{} & \textbf{} & \textbf{} & \textbf{estimation} & \textbf{} \\
			\hline
			\cite{cipriano2012cost} & Compartmental & M1, M2 & Compartmental & US    & S     & M     & No     & L     & E,H,C \\
			\hline
			&       &       & Not   & UK, Canada, &       &       &       &       &  \\
			\cite{martin2013hepatitis} & Compartmental & M1    & incorporated & Australia & M     & S     & No     & Not carried out & E \\
			\hline
			\cite{bennett2015assessing} & Compartmental & M1    & DTMC  & UK    & M     & S     & No     & L     & E,H,C \\
			\hline
			\cite{cousien2016hepatitis} & Agent-based & M1    & DTMC  & France & M     & S     & No     & Not carried out & E \\
			\hline
			\cite{he2016prevention} & Agent-based & M1, M3 & DTMC  & US    & M     & S     & No     & F     & E,H,C \\
			\hline
			\cite{scott2016treatment} & Compartmental & M1    & DTMC  & Australia & M     & S     & No     & W     & H,C \\
			\hline
			\cite{van2016cost} & Agent-based & M1    & DTMC  & Netherlands & S     & S     & No     & W     & E,H,C \\
			\hline
			&       &       & Not   &       &       &       &       &       &  \\
			\cite{ayoub2017impact} & Compartmental & M4    & incorporated & Egypt & M     & S     & No     & Not carried out & E \\
			\hline
			\cite{bennett2017hepatitis} & Compartmental & M1    & DTMC  & UK    & M     & S     & No     & L     & H,C \\
			\hline
			\cite{cousien2018effectiveness} & Agent-based & M1    & DTMC  & France & S     & S     & No     & L     & E,H,C \\
			\hline
			\cite{ayer2019prioritizing} & Agent-based & M1, M3 & DTMC  & USA   & M     & S     & No     & F     & E,H \\
			\hline
			\cite{lim2020effects} & Compartmental & M1, M5, M6 & Compartmental & Pakistan & S     & M     & No     & W     & E/C \\
			\hline
			\cite{kwon2021hepatitis} & Compartmental & M1    & DTMC  & Australia & M     & S     & No     & F     & E,H,C \\
			\hline
			& Dynamic  &       &       &       &       &       &       &       &  \\
			\cite{epstein2023microsimulation} & microsimulation & M1, M7 & DTMC  & USA   & S     & M     & No     & W     & E,H,C \\
			\hline
			This study & Agent-based & M1, M5 & DTMC  & India   & M  & M & Yes     & L & E,H,C \\
			\hline
		\end{tabular}%
	}
	\label{tabstudies}%
\end{table}%

\subsection{Literature Pertaining to \textit{RO1}: Model-based Cost-effectiveness Analyses of HCV Treatment}
\label{subsec2.2}
The literature relevant to \textit{RO1} pertains to model-based cost-effectiveness analyses of HCV treatment - in particular, treatment with DAAs.

Studies that involve the use of dynamic models for assessing the impact of DAAs on health and economic outcomes are the most relevant to this study. \cite{he2016prevention} developed an ABM for HCV transmission through injecting drug use, and then assumed that a certain proportion of transmission occurs via an aggregate mode referred to as `other than injecting drug use'. This model was then used to evaluate the cost-effectiveness of risk-based and universal opt-out screening policies in prisons. This simulation model was later used by \cite{ayer2019prioritizing} to test index-based policies, developed through restless bandit modeling, for prioritizing treatment for HCV patients in prisons. \cite{ayoub2017impact} studied the impact of treatment on HCV epidemiology and health economic outcomes via a deterministic aggregate compartmental model set in Egypt, and considered injecting drug use and blood transfusions as transmission modes. More recently, \cite{epstein2023microsimulation} employed an individual-level modeling approach for HCV transmission among a fixed cohort of hemodialysis patients wherein HCV infections are introduced in dialysis centers via injecting drug use among patients. %\cite{cipriano2012cost} used a deterministic aggregate compartmental model incorporating transmission through injecting drug use and the sexual route to study the CE of screening strategies to identify HCV and HIV among IDUs.

The only dynamic model set in South Asia, in our knowledge, is described in \cite{lim2020effects}: a compartmental model set in Pakistan to model creation of new HCV infections through not only injecting drug use, but also via medical procedures and tattooing. Being an aggregate model, it did not explicitly model individual-level interactions for disease transmissions. In the Indian context, aside from preliminary work underpinning this study \citep{das2019abm}, the relevant studies are all static models that do not consider HCV transmission \citep{aggarwal2017cost,goel2018cost,chugh2019real,chugh2021cost}. These studies incorporate a disease progression discrete-time Markov chain (DTMC) that simulates the progression of the disease to evaluate the impact of different DAA regimes on the health and cost outcomes. Given that the models did not incorporate the creation of new infections, the secondary benefits of treatment in preventing new infections could not be estimated.%\cite{lim2020effects} used this model to explore the impact of multiple treatment strategies using DAAs. 

Only a few studies, in our knowledge, have quantified the effects of stepping up treatment uptake on cost and health outcomes \citep{he2016prevention,bennett2017hepatitis,ayer2019prioritizing,lim2020effects}. In fact, we identified only one study that addressed a question relevant to \textit{RO1}: quantify the impact of not considering transmissions on the cost-effectiveness of DAAs: \cite{bennett2017hepatitis}. The authors used a deterministic compartmental model to capture HCV transmission in a population of IDUs. The study reports the impact of neglecting transmissions on treatment cost-effectiveness at multiple treatment uptake rates. The authors found that the cost-effectiveness of adopting DAAs over previous treatment regimens with lower cure rates improved with increasing uptake rate when transmissions are included. In the Indian context, the studies of \cite{chugh2019real} and \cite{chugh2021cost} explored the effects of increasing treatment uptake rates on DAA cost-effectiveness, but with static DTMC based models. With respect to these studies, and in particular \cite{bennett2017hepatitis}, we see that dynamic individual-level models that consider the inherent stochasticity in disease transmission and spread, and therefore also characterize the uncertainty around their estimated outcomes, have not been developed: (a) for the South Asian or Indian contexts and more broadly for the developing world; (b) that consider modes of transmission other than injecting drug use; (c) and to answer the research questions we consider as part of \textit{RO1} (mentioned previously in Section~\ref{intro}). Thus our research contributions pertaining to \textit{RO1} involve addressing the above gaps.

\subsection{Literature Pertaining to \textit{RO2}: Impact of Intervention Frequency and Timing}
\label{ro2lit}

We found limited work that assesses the cost-effectiveness of changing frequency and timing of interventions for HCV. This appears to be the case for other infectious diseases as well, with relevant studies being conducted in deterministic settings - for example, \cite{duijzer2018benefits} for influenza via a compartmental model, and \cite{deo2015planning} for HIV via a hybrid compartmental and integer programming model. 

For HCV, \cite{cipriano2012cost} evaluates screening strategies among injecting drug users undergoing opioid replacement therapy. The authors found that more frequent screening strategies are more cost-effective. Contrastingly, \cite{epstein2023microsimulation} studied multiple screening frequencies on individuals undergoing hemodialysis, and reported that lower screening frequencies were more cost-effective. Both studies only modelled creation of new HCV infections through the injecting drug use route. The studies also did not analyze the epidemiological, health and economic impact of timing of screening and/or treatment during the study time horizon. 

Therefore, our research contribution with respect to \textit{RO2} involves estimating the epidemiological, health and economic impact of \textit{both} the timing and frequency of HCV treatment with DAAs as a function of treatment uptake rate, accomplished via an ABM developed for the South Asian and developing world contexts that explicitly incorporates HCV transmission via modes other than injecting drug use. Importantly, our study also highlights the trade-off between the `optimal' policy in terms of its cost-effectiveness and its epidemiological effectiveness in terms of the number of infections created. 

 %Thus our research contribution pertaining to \textit{RO3} involves the development of this approach, which addresses the challenges in estimating lifetime outcomes for agents of interest within an ABM (mentioned above) in a relatively more computationally efficient manner.

\section{Agent-Based Simulation Model Development}
\label{sec3}
The ABM that we develop to accomplish research objectives 1 and 2 consists of the following components, each of which we describe in this section.

\begin{enumerate}
    \item Agent interaction environments and ancillary dynamics (e.g., agent demographic model, model run settings).
    \item Disease progression module. 
    \item Treatment module.   
    \item Outcomes estimation module.
\end{enumerate}

%We describe each of the above components now. 

\subsection{Agent Interaction Environments and Ancillary Dynamics}
\label{sim:meth}
Two agent interaction environments were incorporated into the model, corresponding to the dominant modes of transmission in the Indian context: a medical environment and a social interaction environment. The medical environment was chosen based on a study by \cite{chakravarti2013study} that reported the key modes of transmission for HCV in the Indian context. The authors found that 74.1\% of HCV infections were caused via unsafe medical procedures, which included blood transfusions, injections, dental procedures and surgeries. Injecting drug use was the next major mode of transmission, accounting for nearly 16\% of infections, which we incorporated as part of the social interaction environment. \cite{chakravarti2013study} also found that 9\% of infections were caused via unsafe tattooing practices. However, we were unable to obtain sufficiently reliable publicly available information regarding this mode of transmission, and hence we did not explicitly incorporate this environment. Instead, given that the spread of HCV via tattooing is also due to needle-stick injuries, similar to injecting drug use, we modified certain parameter estimates associated with the social interaction environment to incorporate this transmission mode.

In addition, a dynamic agent cohort is incorporated, meaning agents enter and exit the model via birth and death during the simulation execution period. The simulation execution period was divided into two parts: the first 50 years of simulation time, called the calibration period, and then 10 years where the intervention (different treatment models with DAAs) was introduced and outcomes of interest were collected for analysis. The calibration period ensured that at the point in the simulation execution period when collection of outcomes of interest begins, demographics of the agent cohort and the spread of HCV in terms of its prevalence matched those documented in the literature relevant for validating the model. A 10 year intervention period was chosen as a time scale sufficient to capture impacts of different interventions at a population level. Further, daily time points were used so that interactions relevant to the spread of the disease could be captured with the appropriate granularity. %More details regarding validation are provided in \ref{appvalid}. %, as indicated in the literature \citep{edejer2003making}

The model was initialized with 40,000 agents at the beginning of the calibraton period. Among this cohort, 15 IDUs and 5 infected agents were included. Only agents between the ages of 23 and 102 were included. These demographic attributes were chosen for initializing the agent cohort based on preliminary experiments conducted to ensure that the required agent cohort demographics and HCV spread were achieved at the end of the calibration period. Average annual birth and death rates of 15 and 6 per 1,000 persons, respectively, were used in the model, with the death rates decomposed into lower and higher death rates for older and younger agents, respectively. 

\subsubsection{Medical environment for HCV transmission.}
\label{medtrans}

\begin{figure}[htp]
\begin{subfigure}{\textwidth}
\includegraphics[width=\textwidth,height=5cm]{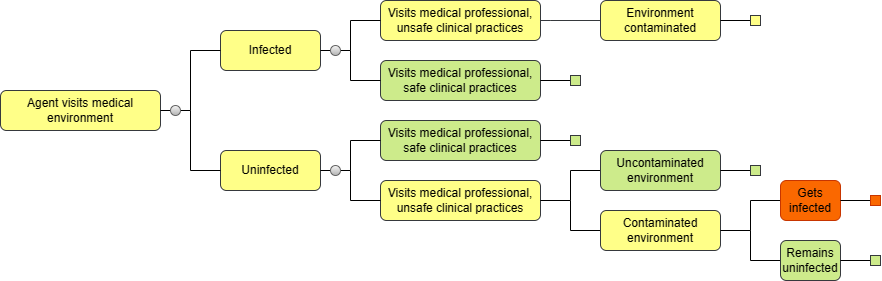}
\caption{Agent interactions in the medical environment.}
\label{fig:med}
\end{subfigure}

\bigskip

\begin{subfigure}{\textwidth}
\includegraphics[width=\textwidth,height=5cm]{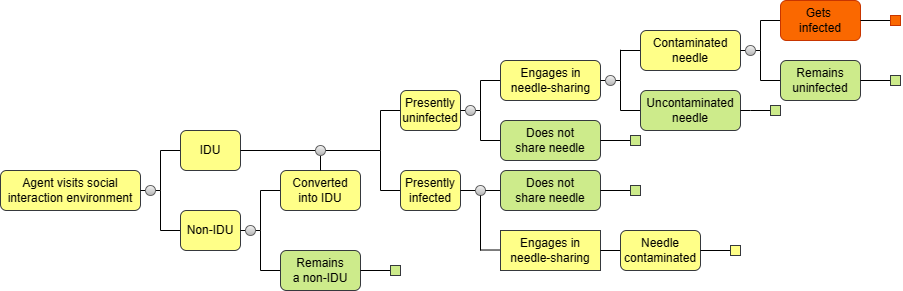}
\caption{Agent interactions in the social interaction environment.}
\label{fig:soc}
\end{subfigure}
\caption{Agent interaction models in the HCV dynamic transmission model.}
\label{fig:env}
\end{figure}

In this environment, two types of agents interact: medical professionals and `normal' agents from the general population. On a given day, each agent may visit the medical environment with a certain probability, where they interact with a medical professional who, again with a certain probability, may follow unsafe clinical practices. If an infected agent visits such a medical professional, then that particular professional's environment gets contaminated. Thus, other uninfected agents who visit that medical professional on that day may get infected. This interaction model is depicted in Figure~\ref{fig:med}.

The above interaction model needs to be parameterized with the appropriate agent-specific probabilities. This includes, for example, the daily probability of an agent visiting the medical environment, the probability that a medical professional employs unsafe clinical practices, and the probability that an uninfected agent gets infected from a contaminated medical environment. %The estimates and sources for all model parameters are given in Table \ref{tabmed}

The daily probability of an agent visiting the medical environment (denoted by $q_v^{med}$) was estimated via the annual per-capita average numbers of the relevant procedures in the Indian context - injections, surgeries, dental procedures and blood transfusions. $q_v^{med}$ is estimated via equation~\ref{eqint} below.
\begin{equation}
\label{eqint}
    q_{v}^{med} =   \frac{m_i + m_b + m_s + m_d}{360} \\
\end{equation}

In equation~\ref{eqint}, $m_j$ denotes the annual average number of times the $j^{th}$ medical event of interest occurs, where $j$ $\in$ \{injections ($i$), blood transfusions \hspace{2pt} ($b$), surgeries ($s$), dental procedures ($d$)\}. The proportion of medical procedures that were unsafe in the Indian context ($q_{uns}^{med}$) was estimated from \cite{pala2016assessment}, which indicated that nearly half the clinics across the country did not follow safe sterilization practices, and thus $q_{uns}^{med}$ was set to 0.50. Next, the number of agents seen by a medical professional on a given day, denoted by $N_{p|mp}$, was estimated using the doctor-patient ratio ($r_{d|p}$), reported to be 1:1800 (see Table~\ref{tabmed}). We thus derived $N_{p|mp}$ using the relation $N_{p|mp} = \frac{q_{v}^{med}}{r_{d|p}}$. The number of medical professionals was increased with time in accordance with the increase in population. 

The probability of an uninfected agent acquiring an infection from a contaminated medical environment, denoted by $q^t_{med}$, was calculated as a weighted average of the respective probabilities of getting infected via a contaminated version of each of the medical procedures considered in the medical environment.
\begin{equation}
\label{eqpeip}
   q_{t}^{med} =  \frac{\sum_{j\in\{i,b,s,d\}} m_j \times r_j}{\sum_{j\in\{i,b,s,d\}} m_j}
\end{equation}

In equation~\ref{eqpeip}, $r_i$, $r_b$, $r_s$ and $r_d$ the per-event probabilities of contracting an HCV infection due to contaminated injections, blood transfusions, surgeries and dental procedures. Thus, in order to estimate $q_{t}^{med}$, we need estimates for each $r_j$. Estimates for $r_i$ and $r_s$ were available; however, those for $r_b$ and $r_d$ are not. However, \cite{chakravarti2013study} reported that blood transfusions alone accounted for 54\% of the HCV load, and thus formed 74\% of the contribution of the medical environment. We can thus write:
\begin{equation}
\label{eqpbt}
\begin{aligned}
    &m_{b} \times r_{b} = 0.74 \times \sum_{j\in\{i,b,s,d\}} m_j \times r_j \text{ and}\\
    &q_{t}^{med} = \frac{m_{b} \times r_{b}}{0.74 \times (\sum_{j\in\{i,b,s,d\}} m_j}
    \end{aligned}
\end{equation}
Thus only an estimate of $r_b$ is required to estimate $q_t^{med}$. Given that such an estimate was not available from the relevant clinical literature, $q_t^{med}$ was treated as a `calibration' variable, and estimated via the model calibration process (described in section \ref{appvalid}).

\subsubsection{Social interaction environment.}
\label{soctrans}
The social interaction environment was created to incorporate HCV transmissions via injecting drug use. Two key events occur here: a non-IDU can get converted into an IDU, and/or an uninfected IDU gets infected via a needle-sharing event. The interaction model for this environment is depicted in Figure~\ref{fig:soc}. If an uninfected non-IDU agent visits the environment, they may get converted into an IDU. If an IDU visits the environment, then they may engage in a needle-sharing event. If an infected IDU is part of the needle-sharing group then we assume that all IDUs that are part of said group are exposed, and may acquire HCV with a certain probability.

Much of the data required to parameterize the above interaction model was obtained from a study by \cite{ambekar2008size} of IDU characteristics in the Indian states of Punjab and Haryana. We assume that a non-IDU visits this environment once a week. The daily probability of an IDU visiting the social interaction environment, denoted by $q_v^{soc,I}$ was estimated using the weekly frequency (e.g., twice a week, thrice a week) of injecting observed among IDUs in the Punjab state \citep{ambekar2008size}. This was calculated as the weighted average of the injecting frequencies observed in the districts of Punjab, with the weights being the populations of the individual districts. Further, only agents between the ages of 16 and 32 were eligible to become IDUs, based on the observation by \cite{ambekar2008size} that 75\% of IDUs were aged between 18 and 30 years. Similarly, the authors also reported that duration for which agents remained as IDUs was 3 years. The size of an IDU group that may engage in a communal injecting event wherein needles may be shared was determined to be 3, based on a study by \cite{azim2008hiv} in Bangladesh (similar data was not available for the Indian context). Note that an overestimation of the average IDU group size reported in \cite{azim2008hiv} was made to account for HCV transmission via tattooing, as discussed earlier.

The probability of a non-IDU getting converted into an IDU is estimated by conditioning on whether an agent is unemployed or not. This is because the rate of unemployment among IDUs was observed to be 26\%, compared to 16.6\% among the general population \citep{ambekar2008size,mospi23}. We developed the following model (equation~\ref{eq:iduconv}) to incorporate the association of unemployment with the probability of a non-IDU being converted into an IDU (denoted by $q_{c}$).
\begin{equation}
\label{eq:iduconv}
\centering
q_{c} = q_{c}^{ue} \times q^g_{ue} + q^{e}_{c} \times (1 - q^g_{ue}) 
\end{equation}

Here $q_{c}^{ue}$ and $q_{c}^{e}$ represent the probabilities of an unemployed non-IDU agent versus that of an employed agent getting converted into an IDU. The odds of an unemployed non-IDU getting converted into an IDU as compared to an employed non-IDU was taken as the ratio of the probability of being unemployed for an IDU ($q_{ue}^{I} = 1 - q_{ue}^g$) to the probability of being unemployed for the general population ($q_{ue}^g$). Thus, we have: $\frac{q_{c}^{ue}}{q_{c}^{e}} = \frac{(1 - q_{ue}^{g})}{q_{ue}^{g}}$. Estimates for $q_{ue}^{I}$ and $q_{ue}^{g}$ were available, but the values of $q_{c}$, $q_{c}^{ue}$ and $q_{c}^{e}$ were unknown, yielding three unknown variables and two equations. Therefore, $q_{c}$ was also assumed to be a calibration variable. Once its estimate was obtained using calibration, the values of $q_{c}^{ue}$ and $q_{c}^{e}$ were found by applying the above relationships in equation~\ref{eq:iduconv}. 

The probability of having shared needles at least once was reported to be 0.504 by \cite{ambekar2008size}. Given an event of the IDU visiting the social interaction on a particular day, we denote the per-event probability of a needle-sharing event among the IDU's group as $q_{shar}$. However, we could not identify an estimate for $q_{shar}$ from the literature, and hence this also was estimated via model calibration. However, an upper bound of 0.504 guided the calibration process for this parameter.

Finally, the probability of acquiring the HCV infection from using a contaminated needle - i.e., as part of a needle-sharing injecting event among a group of IDUs (of size 3, as discussed earlier) with at least one infected agent - was estimated as 0.02 \citep{rolls2012modelling}.

The agent-based model input parameters associated with the medical and social interaction environments are provided in Table~\ref{tabmed} below. 

\begin{table}[htbp]
	\centering
	\caption{Input parameters for the medical and social interaction environments of the HCV agent-based model.}
	\resizebox{\textwidth}{!}{
		\begin{tabular}{|r|cc|c|c|}
			\hline
			\multicolumn{1}{|c|}{\textbf{Environment}} & \multicolumn{2}{c|}{\textbf{Relevant parameter for HCV transmission}} & \textbf{Value} & \multicolumn{1}{c|}{\textbf{Source}} \\
			\hline
			& \multicolumn{1}{c|}{} & Injections, $m_i$ & 2.9   & \cite{ipen2012injection} \\
			\cline{3-5}          & \multicolumn{1}{c|}{Annual per-capita} & Blood transfusions, $m_b$ & 0.023 & \cite{naco18} \\
			\cline{3-5}          & \multicolumn{1}{c|}{average number of} & Surgeries, $m_s$ & 0.009 & \cite{weiser2016size} \\
			\cline{3-5}    & \multicolumn{1}{r|}{} & Dental procedures, $m_d$ & 0.982 & \cite{sta18} \\
			\cline{2-5}          & \multicolumn{2}{c|}{Daily probability of an agent} &   & Calculated using equation~\ref{eqint}  \\
                      & \multicolumn{2}{c|}{visiting the medical environment, $q_{v}^{med}$} & 0.0108 & (section \ref{medtrans}) \\
			\cline{2-5}    \multicolumn{1}{|c|}{Medical}      & \multicolumn{2}{c|}{Doctor to patient ratio, $r_{d \mid p}$} & 1:1800  & \cite{deo2013doctor} \\
			\cline{2-5}     \multicolumn{1}{|c|}{environment}      & \multicolumn{2}{c|}{Number of patients per medical professional, $N_{p \mid mp}$} & 19  & Calculated as $\frac{q_{v}^{med}}{r_{d \mid p}}$ \\
			\cline{2-5}          & \multicolumn{2}{c|}{Proportion of medical professionals following unsafe practices, $q_{uns}^{med}$} & 0.50  & \cite{pala2016assessment} \\
			\cline{2-5}          & \multicolumn{2}{c|}{Contribution of medical procedures to HCV infection load} & 74.10\% & \cite{chakravarti2013study} \\
			\cline{2-5} & \multicolumn{2}{c|}{Per-event probability of infection from blood transfusion, $r_b$} & 0.806 & Calibrated \\
			\cline{2-5} & \multicolumn{2}{c|}{Per-event probability of infection in the medical environment, $q_t^{med}$} & 0.0064 & Calibrated \\
			\hline
			& \multicolumn{2}{c|}{Proportion of IDUs in the age range of 18-30 years} & 75\%  & \cite{ambekar2008size} \\
			\cline{2-5}          & \multicolumn{2}{c|}{Age range of IDUs in the model} & 16-32 years & Age-wise uniformly distributed \\
			\cline{2-5}          & \multicolumn{2}{c|}{Daily probability of an IDU visiting the environment} & 0.486 & \cite{ambekar2008size} \\
			\cline{2-5}     & \multicolumn{2}{c|}{Daily probability of a non-IDU visiting the environment} & 0.142 & Weekly visits assumed \\
			\cline{2-5}          & \multicolumn{2}{c|}{Probability of an IDU being unemployed, $q_{ue}^I$} & 0.26  & \cite{ambekar2008size} \\
			\cline{2-5}          & \multicolumn{2}{c|}{Probability of a non-IDU agent being unemployed, $q_{ue}^g$} & 0.166 & \cite{indiaspend17} \\
			\cline{2-5}       \multicolumn{1}{|c|}{Social}   & \multicolumn{1}{c|}{Probability of conversion of a non-IDU} & employed agent, $q_{c}^{e}$ & 1.44 $\times$ $10^{-5}$   & Calibrated \\
			\cline{3-5}    \multicolumn{1}{|c|}{interaction}      & \multicolumn{1}{c|}{into an IDU for an} & unemployed agent, $q_{c}^{e}$ & 2.25 $\times$ $10^{-5}$ & Calibrated \\
			\cline{2-5}
			\multicolumn{1}{|c|}{environment}   & \multicolumn{2}{c|}{Duration for which an IDU remains an IDU} & 3 years & \cite{ambekar2008size} \\
			\cline{2-5}          & \multicolumn{2}{c|}{Size of an IDU network} & 3 persons & Based on \cite{azim2008hiv} \\
			\cline{2-5}         
			& \multicolumn{2}{c|}{Probability of an IDU having shared a needle at least once} & 0.504 & \cite{ambekar2008size} \\
			\cline{2-5}    \multicolumn{1}{|c|}{} & \multicolumn{2}{c|}{Per-event needle sharing probability in a network of size 3, $q_{shar}$} & 0.41 & Calibrated \\
			\cline{2-5}          & \multicolumn{2}{c|}{Per-event probability of getting} &       &  \\
			& \multicolumn{2}{c|}{infected due to sharing needles, $q_t^{shar}$} & 0.02  & \cite{rolls2012modelling} \\
			\hline
		\end{tabular}%
	}
	\label{tabmed}%
\end{table}%

With regard to the disease progression module of the ABM, we use a validated DTMC for modelling HCV progression that has been widely used in cost-effectiveness studies of interventions for HCV \citep{ayer2019prioritizing}. This model has been used in all the studies which evaluated the cost-effectiveness of DAAs in the Indian context \citep{aggarwal2017cost,goel2018cost,chugh2019real,chugh2021cost}. We refer to this model henceforth as the NH-DTMC, and its implementation within the ABM for each infected agent can be considered as the execution of a Monte Carlo simulation replication. A detailed description of the NH-DTMC and its implementation for each infected agent (Algorithm~\ref{algupdt}) is provided in Section \ref{appdisprog}.

%\subsection{Disease Progression Module for Infected Agents}
%\label{sec:proghcv}
 
 \subsection{Treatment Module for Infected Agents}
 \label{treatmod}
Treatment was considered only for agents for whom treatment had not failed earlier given that with DAAs, the probability of treatment failure is low \citep{chugh2019real}. DAAs are the current standard of care in India, as in much of the rest of the world. \cite{chugh2019real} estimated real-world cure or SVR rates for DAAs from their experience of treating HCV patients in Indian Punjab. The authors found that the pan-genotypic combination treatment of DAAs sofosbuvir (SOF) and velpatasvir (VEL) yielded the highest SVR rates, and hence we used SOF+VEL as the standard of care in our simulation experiments, with genotype-specific SVR rates (ranging from 84\% to 86\%) sourced from \cite{chugh2019real}.%genotype distribution sourced from \cite{mehta2018impact} and

%The SVR rates vary according to the disease genotype and disease state \citep{chugh2019real}. The distribution of genotypes was taken from \cite{mehta2018impact}, conducted in the Indian province of Punjab. The study found that 72\% of HCV patients were of genotype G3, 22\% were of genotype G1 and 6\% were of genotype G4. Thus, when an agent is infected, we assign the genotype of their HCV infection based on the above probabilities, and then apply genotype-specific SVR rates from treatment (if they are treated). \cite{chugh2019real} was used as a source for real-world genotype-specific SVR rates in the Indian context. From this study, we found that for treatment of non-cirrhotic HCV patients (in states F0 to F3), the SVR rate was 86\% for all genotypes. Hence, this SVR rate was applied to all infected agents if they received treatment in stages (NH-DTMC states) F0 to F3. For agents in the F4 state, an SVR rate was 84\% for those with a genotype G3 infection, and 86\% for those with infections of genotypes G1 and G4. For DC patients, an SVR rate was 84\% was applied for all genotypes. The genotype of the HCV infection carried by an agent is relevant only from a treatment standpoint, given that the NH-DTMC used to model disease progression is pan-genotypic. 

We now discuss how the treatment is applied to the agent cohort in the intervention period. First, we recall that the simulation execution period is divided into two stages: a calibration period of 50 years of simulation time during which the demographics of the agent cohort and the spread of the HCV infection reach levels corresponding to the validation targets (published real-world observations of HCV prevalence). A 10-year intervention period then begins during which treatment using DAAs is applied to a subset of infected agents and outcomes pertaining to HCV epidemiology and its health and cost impacts are recorded.

In this study, we evaluate multiple treatment models or approaches towards the treatment and management of HCV. These involve varying the timing and frequency of treatment - for example, running treatment camps once every 3 years, or at the beginning of the intervention period versus the end of the intervention period. Each of these alternate treatment models are compared against the default model that represents the status quo and is referred to henceforth as the \textit{annual-treatment} model. This treatment model, under which an increasing proportion of infected agents are treated every year (hence the name), represents the status quo in the following way. We assume that, based on inputs from our clinical collaborator with field experience in managing HCV treatment in a high-prevalence area, beginning from a base value of the proportion of the infected cohort who receive treatment (capturing the impact of initial awareness regarding HCV), public awareness regarding HCV increases every year. The number of infected agents who receive treatment via implementation of the default \textit{annual-treatment} model over the intervention period is controlled by the target \textit{uptake rate}. For example, if the target uptake rate is set to be 10\%, then our implementation of \textit{annual-treatment} attempts to ensure that 10\% of all infected agents present from the beginning of the intervention period (including those infected during the calibration period) to the end of the intervention period are treated. 

We now discuss the approach towards implementing the \textit{annual-treatment} model so that the target uptake rate is achieved. Given that some infected agents die before undergoing treatment, hence the target uptake rate is not precisely achieved during a given simulation run. Further, the naive approach of treating 10\% of infected agents every time period (e.g., a year) would not capture the scenario of increasing numbers of infected agents receiving treatment every year. Therefore, we designed an algorithm that, starting from a base proportion of infections treated, takes into account the new infections created every year and steps up the proportion treated every year to achieve the target uptake rate. The algorithm that we design for this purpose may be more broadly applicable: it can be used, in a dynamic transmission model such as an ABM or an ordinary differential equation based compartmental model, to implement any intervention that needs to be gradually stepped up over regular time points to achieve a preset coverage rate. In a compartmental model, our algorithm can be used with a constant intervention effectiveness rate applied to the relevant compartment (e.g., the `infected' compartment in an SEIR model) within the model. We carried out all our simulation experiments on treatment of HCV using SOF+VEL at six target treatment uptake rates- 10\%, 30\%, 50\%, 70\%, 90\% and 95\%. 

 The implementation of the annual treatment model is described in Algorithm~\ref{alganntrt} below. In this model, treatment takes place at discrete time points across a time horizon (e.g., every day over a period of one week or in our case, every year across a 10-year intervention period). We assume that new infections take place at the beginning of the time point, then treatment takes place for a certain proportion of the then (living) infected population. After treatment, the remaining infections result in new infections in the following time point. We do not consider non-disease-related deaths while outlining the algorithm for the sake of simplicity. 

\noindent \underline{\textit{Notation}}
%\begin{itemize}

-- $j$, $i$ = Time indices.

%\item[--] $ag_{j}^{new}$ = An agent that gets infected at time point $j$.

-- $N_j^{new}$ = Number of new infections caused at time point $j$.

-- $N_j^{a}$ = Number of new infections caused at time point $j$ which are assigned for treatment.

-- $N_{j_1,j_2}^{Tr}$ = Number of new infections caused at time point $j_1$ which are treated at time point $j_2$, where 0 $\leq$ $j_1$ $\leq$ $j_2$ $\leq$ $n$.

-- $M_j^{Tr}$ = Number of infections treated at time point $j$. 

-- $M_j^l$ = Number of infections assigned to treatment remaining untreated after treatment has concluded at time point $j$.

-- $n$ = Number of time points in the intervention period.

-- $N^0$ = Number of initial infections, or the number of infections at the beginning of the intervention period.

-- $u \in [0,1]$ = Target uptake rate of the treatment or the proportion of patients assigned to treatment. 

-- $\alpha \in [0,1]$ =  Initial coverage rate or proportion of the initial pool of infected patients assigned to treatment who undergo treatment at the beginning of the intervention period.

%\end{itemize}

\begin{algorithm}
	\caption{Implementation of the annual-treatment model.}
	\label{alganntrt}
	\begin{algorithmic}
		\For{$j = 0 \to n$}
		\If{$j = 0$}
		\State $N^{new}_j = N^0$
		\Else
		\State $N^a_j = u \times N^{new}_j$
		\If{$j = 0$}
		\State $M^{Tr}_j = \alpha \times N^a_j$
		\Else
		\State $M^{Tr}_j = \left[ \alpha + \left(1 - \alpha \right) \times \frac{j}{n} \right] \times \left( M_{j-1}^l + N^a_j \right)$
		\State where $M_{j-1}^l = \sum_{i=0}^{j-1} N^a_i - \sum_{i=0}^{j-1} M^{Tr}_i$
		\EndIf
		\EndIf
		\EndFor
	\end{algorithmic}
\end{algorithm}

In Algorithm~\ref{alganntrt}, $j$ = 0 is the beginning of the period and $j$ = $n$ is the end of the intervention period, with treatment occurring over $n + 1$ time periods. If $u$ is the target uptake rate across the intervention period, then we assume that a fraction $\alpha$ of $u$ (e.g., 30\% of all those to be treated, which may be, say, 10\% of all infected agents present across the intervention period) are treated at the end of $j = 0$. The coverage of treatment in each time period is then scaled upwards linearly to reflect the likelihood of awareness increasing with time. In each time period $j > 0$, we assign $u$\% of newly created infections to be treated ($N^a_j)$, and then set the number treated in $j$ to the coverage fraction $\left[\alpha + (1 - \alpha)\times \frac{j}{n}\right]$ multiplied by the total number of infections present at $j$, which is the sum of the number of infections remaining at the end of $j - 1~(M^l_{j-1}$) and $N^a_j$. Note that if $\alpha \times u$ fraction of the infections present prior to $j = 1$ are treated at $j = 0$, then scaling up the coverage linearly implies that with each time period, a fraction of the remaining coverage (equal to $1 - \alpha$) to be achieved must be added to the base coverage at $j = 0$, $\alpha$. To ensure that coverage at $j = n$ is equal to $u$ (or in other words, $\alpha = 1$), this fraction is set to $\frac{j}{n} \times (1 - \alpha)$.

Due to page length constraints, further details of the annual treatment model is described in detail in Section \ref{aptrt}. There, we provide a numerical illustration of the implementation of Algorithm~\ref{aptrt}, as well as a straightforward analytical argument to show that it achieves the desired uptake rate without non-disease-related mortality.

%In \ref{aptrt}, we provide a numerical illustration to illustrate the implementation of the \textit{annual-treatment} algorithm, and also a straightforward mathematical argument to show that Algorithm~\ref{alganntrt} yields the desired uptake rate under deterministic conditions and without non-disease-related deaths.

\subsection{Outcomes Estimation Module}
\label{outcomes}

Prior to describing the estimation of outcomes, we point the reader to section \ref{appvalid}, where we describe the validation of the simulation model.

\subsubsection{Outcomes of interest.}
\label{sec:outint}

The outcomes generated by the model are of three types: epidemiological outcomes such as HCV antibody and RNA prevalences, as discussed in section \ref{appvalid}; health outcomes such as life-years (LYs) and quality-adjusted life-years (QALYs); and economic outcomes such as the average total costs of HCV incurred per agent (infected and uninfected - these are \textit{population-level} outcomes) and the average total costs of HCV incurred per chronically infected patient (\textit{patient-level} outcomes). In addition, a summary measure of the cost-effectiveness of the intervention under consideration called the net monetary benefit (NMB) is also generated as a function of the above health and economic outcomes.

The generation and validation of epidemiological outcomes is discussed in section \ref{appvalid}. We now discuss the generation of \textit{lifetime horizon} health and economic outcomes, and our novel approach towards generating these efficiently within a dynamic transmission model such as an ABM. 

Outcomes such as the LYs for each agent are generated by simply recording the time spent by the agent in the model and in each disease state. QALYs associated with each state are estimated by multiplying a utility weight capturing the health-related quality of life associated with the health state (ranging between 0.0 for death and 1.0 for perfect health) with the time spent in the state. Summing this product across all states (including the uninfected state) yields the QALYs associated with each agent in the model. HCV health state costs are similarly estimated: the recurring and fixed costs incurred by being in a given health state are aggregated along with the resource use (e.g., the frequency of resource use - weekly, monthly) and time spent in the state to yield the costs associated with the health state. These are then summed across each state to yield the total costs associated with the management of HCV for an infected agent. We note here that only the direct medical costs associated with managing HCV are incorporated in our analysis, such as the costs of treatment, physician visits, diagnostics and monitoring, surgeries and hospitalizations.

The benchmark approach towards estimating LYs, QALYs and costs for each infected agent would involve executing the NH-DTMC and the non-disease-related mortality (non-DRD) models every time step, recording the LYs (length of one time step - e.g., one day or year), QALYs and costs incurred in the time step, and adding it to the outcomes accumulated until the previous time step. This process would have to be repeated until the agent dies or reaches the maximum permissible age. This termination time point may extend well beyond the ABM execution time horizon. We refer to this process as the \textit{incremental accumulation} (IA) outcomes estimation process, and depict it in Algorithm~\ref{alginc} below. Note that Algorithm~\ref{alginc} calls Algorithm~\ref{algupdt}.

\noindent \textbf{Notation:}\\
	-- $t$: time index, wherein $t$ may extend beyond the ABS time horizon $T$.\\
	-- $a_t \in \{1, 2, ..., A\}$: age group that the agent belongs to at time $t$. \\
	-- $Inf_t$: infection status of the agent at time $t$; $Inf_t$ = 1 if infected, 0 otherwise. \\
	-- $k_t \in \{1,\dots, K\}$: disease stage/state at $t$. \\
	-- $Q$: outcome of interest. \\
	-- $Q^{I}_t$: Incrementally accumulated value of $Q^I$ until $t$. \\ 
	-- $\Delta Q^I_t$: incremental value of $Q^I$ at time $t$. \\
	-- \textit{Init}: subroutine that assigns initial values of agent $a_t$, $Inf_t$ and $k_t$ for the agent. \\
	-- \textit{Eval}: subroutine that calculates $\Delta Q^I_t$ to be added to $Q^{I}_{t-1}$ based on $a_t$ and $Inf_t \times k_t$. If uninfected, then $Inf_t \times k_t = 0$, and if infected, $Inf_t \times k_t = k_t \in \{1,\dots, K\}$. \\
	-- \textit{Updt}: subroutine that updates $Inf_t$ and $k_t$ based on the transmission and progression modules.

\begin{algorithm}[htbp]
	\caption{The \textit{incremental accumulation} approach for \textit{lifetime-horizon} outcomes estimation.}
	\label{alginc}
	\begin{algorithmic}
		\State $t \gets 0$ \Comment{$t$ is the point of call of this algorithm for an agent.}
		\State $a_0$, $Inf_0$, $k_0 \gets \textit{Init}(a,Inf,k)$
		\State $Q^{I}_0 \gets 0$
		\While{Agent is alive}
		\Comment{Updates occur at the end of $t$}
		\State $t \gets t + 1$
		\State $a_t \gets a_{t-1}$ \Comment{Update age group}
		\State $Inf_t,k_t \gets \textit{Updt}(Inf,k)$ \Comment{Call Algorithm~\ref{algupdt}}
		\State $\Delta Q^I_t \gets \textit{Eval}(a_t,Inf_t \times k_t)$
		\State $Q^{I}_t \gets Q^{I}_{t-1} + \Delta Q^I_t$
		\EndWhile
		\State \textbf{Return} $Q^{I}_t$
	\end{algorithmic}
\end{algorithm}

In order to calculate the above outcomes, we needed estimates of utility weights and HCV-related costs incurred in the Indian context. We obtained utility weights for HCV health states from \cite{chugh2019real}, who used HCV health state utility weights estimated specifically for the Indian context. Distinct from \cite{chugh2019real}, we applied utility weights for the uninfected health state as well in order to capture the deterioration of quality of life with age. In other words, if $u_i \in (0,1)$ is the utility weight for uninfected agents of the $i^{th}$ age group ($i \in I)$, and $u_k \in (0,1) ~ (k \in K)$ is the utility weight for the $k^{th}$ HCV health state, then the utility weight for an infected agent of age group \textit{i} in health state $k$ is $u_i \times u_k$. The age-group specific utility weights are obtained from \cite{he2016prevention}.

The resource use and costs associated with the management of HCV were obtained from our expert clinical collaborator whose team has been involved in the management and surveillance of HCV in Punjab over multiple decades. The utility weights capturing the age-related deterioration in quality of life by age group and HCV disease state are provided in Table~\ref{tabqol} and the HCV resource use and management costs are provided in Table~\ref{tabcost}. Discounting is applied to the health and economic outcomes generated by the model, wherein LYs, QALYs and costs are discounted at 3\% per year, based on the Indian health technology assessment guidelines \citep{prinja2018health}.

\section{Computational Experiments for Research Objectives 1 and 2}
 \label{numres}
In this section, we describe the computational experiments associated with research objectives 1 and 2 and their results. 

For both $RO1$ and $RO2$, the epidemiological outcomes considered were the HCV antibody, HCV $RNA$, and IDU prevalences at the end of the intervention period. Health outcomes considered include life years and QALYs, and economic outcomes include the direct medical costs associated with the management of the HCV infection. Both undiscounted and discounted estimates of health and economic outcomes were generated. Health and economic outcomes were generated at both the \textit{population-level} as well as at a \textit{patient-level}. Population-level outcomes were generated by dividing estimates of health and economic outcomes by the total number of agents in the simulation model. Patient-level outcomes were estimated by considering the number of chronically infected agents in the model.

The cost-effectiveness of an intervention considered in the model (e.g., treatment `camps' every 5 years with an uptake rate of 50\%) was summarized using the net monetary benefit. Given that our model incorporates transmissions, the NMB for a given intervention was calculated for population-level outcomes alone, as it would account for costs and health outcomes gained by preventing transmissions by curing HCV patients. The NMB of an intervention is calculated as the incremental health benefits gained in terms of QALYs, converted to monetary terms, minus the incremental costs incurred when the intervention is implemented. The NMB calculation is provided in equation~\ref{eqnmb} below.
\begin{equation}
\label{eqnmb}
    NMB = k \times (Q_{n} - Q_{s}) - (C_{n} - C_{s})
\end{equation}

In equation~\ref{eqnmb}, $k$ represents the willingness-to-pay threshold for incremental health benefits, commonly estimated as three times the per-capita annual gross domestic product of the country where the analysis is situated \citep{prinja2018health}. $Q_n$ and $Q_s$ represent the QALYs associated with the intervention and the status quo, respectively, and $C_n$ and $C_s$ represent the costs. The per-capita GDP of India was INR 113,967 in 2022-23 \citep{mospi23}. In almost all cases, the `status quo' intervention against which the NMB of an intervention was calculated was the annual treatment model as described in \ref{alganntrt} with a target uptake rate of 10\%.

The computational experiments for \textit{ROs} 1 and 2 were carried out on multiple workstations with varying capacity: workstations with Intel Core-i7 and Xeon processors with clock speeds ranging from 3.2 GHz to 3.4 GHz and random access memories of 16GB, 32GB, and 64GB, respectively. The runtimes per replication ranged from 4.2 hours to 7.8 hours, respectively, for the most and least powerful workstations.

\subsection{Research Objective 1: Quantifying the Impact of Treatment Uptake and Incorporating Transmissions on DAA Cost-effectiveness}
\label{ro1res}

We first investigated the impact of increases in treatment uptake rate, from 10\% to 95\%, on epidemiological, health and economic outcomes associated with HCV treatment with $DAAs$. The results of this experiment are documented in Table~\ref{tabwithtransmissions}. %The results in Table~\ref{tabwithtransmissions} are generated by averaging outcomes across 3 replications. We only conducted 3 replications due to the low variance observed in the results across the 3 replications - for example, the coefficients of variation ($CV$) of all HCV prevalence outcomes as well as health and economic outcomes were less than 5\% - and due to limitations on the available computational infrastructure. The low variance observed is likely due to a sufficiently large number of agents considered in the simulation. 

\begin{table}[htbp]
  \centering
  \caption{Impact of treatment uptake rate on the cost-effectiveness of directly-acting antivirals. Standard deviations in parentheses.}
   \label{tabwithtransmissions}
  \resizebox{\textwidth}{!}{
    \begin{tabular}{|c|cccccc|}
    \hline
    \hline
    Target treatment uptake rate (\%) & 10\%  & 30\%  & 50\%  & 70\%  & 90\% & 95\% \\
     Effective treatment uptake rate (\%) & 9.5 (0.2) & 29 (0.6) & 47.7 (0.2) & 67.05 (0.8) & 84.9 (0.2) & 89.2 (0.5) \\
    \hline
    \hline
    \multicolumn{7}{|c|}{Epidemiological outcomes} \\
    \hline
    \hline
    HCV Antibody prevalence (\%) & 6.2 (0.2) & 5.6 (0.06) & 5.1 (0.07) & 4.6 (0.02) & 4.1 (0.08) & 4.0 (0.04) \\
    HCV $RNA$ prevalence (\%) & 4.4 (0.1) & 3.2 (0.05) & 2.2 (0.08) & 1.4 (0.05) & 0.7 (0.04) & 0.6 (0.01)\\
    IDU prevalence (\%) & 0.08 (0.01) & 0.08 (0.02) &  0.07 (0.01) & 0.05 (0.01) & 0.04 (0.01) & 0.05 (0.003) \\
    \hline
    \hline
    \multicolumn{7}{|c|}{Health and economic outcomes} \\
    \hline
    \hline
    \multicolumn{1}{|c}{\textit{Population-level outcomes}} &       &       &       &       &  & \\
    \hline
    Life years & 70.00 (0.04) & 70.12 (0.03) & 70.19 (0.04) & 70.28 (0.04) & 70.30 (0.05) & 70.33 (0.03) \\
    Life years (discounted) & 27.15 (0.02) & 27.15 (0.01) & 27.15 (0.01) & 27.15 (0.01) & 27.13 (0.01) & 27.13 (0.01) \\
    QALYs (years) & 59.15 (0.03) & 59.36 (0.02) & 59.52 (0.04) & 59.68 (0.03) & 59.76 (0.05) & 59.80 (0.02) \\
    QALYs (years, discounted) & 23.67 (0.01) & 23.73 (0.01) & 23.78 (0.01) & 23.82 (0.01) & 23.84 (0.01) & 23.84 (0.01) \\
    Costs (INR) & 31469 (1009) & 24684 (362) & 18206 (755) & 12700(489) & 8287 (397) & 7223 (176) \\
    Costs (INR, discounted) & 12935 (390) & 10302 (132) & 7899 (282) & 5812 (137) & 4160 (128) & 3780 (43) \\
    Net monetary benefit (NMB) & 0     & 23147 & 42645 & 58408 & 66898 & 67278 \\
    \hline
    \multicolumn{1}{|c}{\textit{Patient-level outcomes}} &       &       &       &       &  & \\
    \hline
    Life years & 71.04 (0.46) & 72.87 (0.43) & 75.10 (0.17) & 77.10 (0.40) & 78.50 (0.32) & 78.95 (0.65) \\
    Life years (discounted) & 39.10 (0.21) & 39.93 (0.19) & 41.07 (0.08) & 42.20 (0.21) & 43.12 (0.33) & 43.37 (0.28) \\
    QALYs & 48.15 (0.35) & 51.58 (0.41) & 55.37 (0.15) & 58.90 (0.27) & 61.76 (0.40) & 62.56 (0.48) \\
    QALYs (discounted) & 27.68 (0.19) & 29.30 (0.19) & 31.15 (0.06) & 32.96 (0.17) & 34.38 (0.32) & 34.88 (0.23) \\
    Costs & 795230 (4699) & 676428 (2696) & 543451 (10942) & 420211 (15172) & 303418 (10759) & 270807 (6330) \\
    Costs (discounted) & 326869 (1227) & 282326 (1172) & 235821 (3261) & 192304 (4052) & 152358 (3167) & 141703 (1614) \\
    \hline
    \hline
    \end{tabular}%
    }%
\end{table}%

With regard to epidemiological outcomes, we observe that, as expected, increasing treatment uptake rate leads to substantial decreases in HCV antibody as well as $RNA$ prevalence. In fact, up to (and including) an uptake rate of 90\% the rate of decrease in these prevalences actually increases with increase in uptake rate, implying epidemiological benefits to be gained with improving uptake rate even for treatment programmes performing reasonably well (e.g., uptake rates around 75\%). We also see these benefits reflected in epidemiological dynamics across the intervention period: for example, the HCV antibody prevalence increased from 3.6\% to 6.2\% at the end of the intervention period with the 10\% uptake rate, whereas it increased from 3.6\% to 4.1\% with the 90\% uptake rate. However, limited epidemiological gains are observed when the uptake rate exceeds 90\% - for example, the decrease in HCV antibody prevalence is not statistically significant when uptake rate increases from 90\% to 95\%. %Similar trends - i.e., changes that are not statistically significant - are also observed for HCV $RNA$ and IDU prevalences.%We also note substantial decreases in the IDU prevalence with increasing uptake rates, up to 90\%. This is because more infected IDUs stop injecting drug use after diagnosis and treatment with increases in uptake rate.

When population-level health and economic outcomes are considered, increasing treatment uptake is cost-effective at all uptake rates exceeding 10\%, as all NMB estimates are positive. However, increasing uptake rate yields only marginal changes in population-level life years and  QALYs. This is not surprising given that: (a) across all the outcomes in Table~\ref{tabwithtransmissions}, the maximum HCV antibody prevalence at the end of the intervention period, which indicates the extent to which the population has been affected, is only 6.2\% (at the lowest uptake rate of 10\%); and (b) even among this small set of infected agents, chronic HCV does not, on average, impact life expectancy. On the other hand, we see significant savings in costs and corresponding increases in the NMB estimates with increasing uptake. The rate of increase in the NMB, however, decreases with increasing uptake. %In fact, the increase in cost savings increases with uptake rate: for example, it is approximately 20.3\% when uptake rate changes from 10\% to 30\% versus 28.4\% when it changes from 70\% to 90\%.%: the NMB increases by 84\% when uptake rate increases from 30\% to 50\% and by approximately 9\% when uptake rate increases from 90\% to 95\%

\begin{figure}[htp]
    \centering
    \includegraphics[width=12cm]{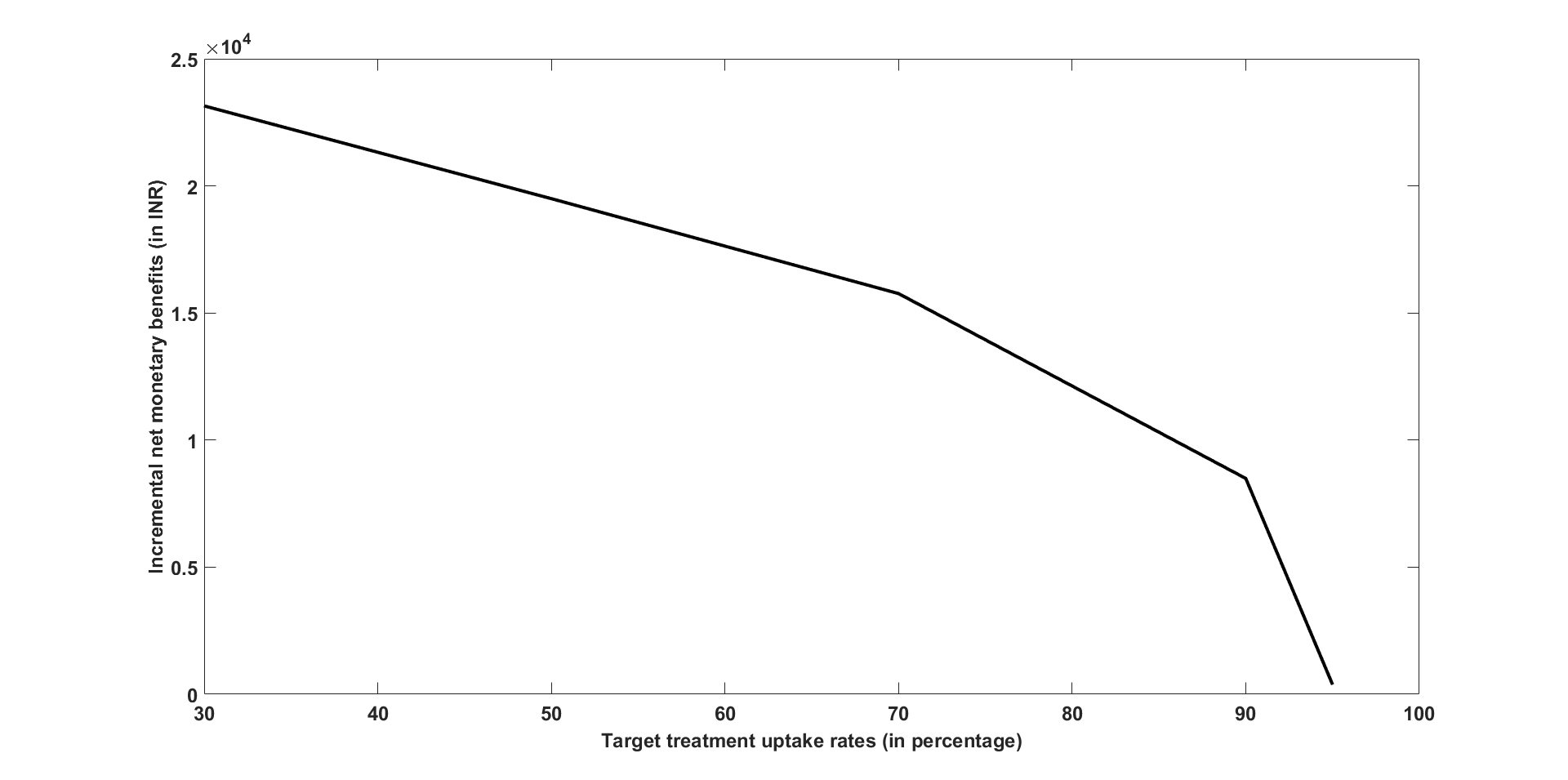}
    \caption{Incremental net monetary benefit at each applicable treatment uptake rate, calculated with respect to the previous uptake rate.}
    \label{figinmb}
\end{figure}

An important question arises here: in resource-constrained settings, how do we quantitatively determine the point beyond which efforts to increase uptake rate may cease? We suggest that this question may be answered by calculating the NMB associated with each uptake rate assuming that the comparator is the largest uptake rate smaller than the uptake rate in question. This is expressed in equation~\ref{eqincnmb} below. 
\begin{equation}
\label{eqincnmb}
    \begin{aligned}
    &NMB(u_i) = k \times (Q(u_i) - Q(u_j)) - (C(u_i) - C(u_j)) ~\forall ~ u_i \in U \\
    &\text{where } U = \{u_1, u_2,\dots,u_K\}, ~u_{j} = \max\{u_k: k \neq i, u_k < u_i\} \\ 
    &\text{ and } u_i \neq \min U \\
\end{aligned}
\end{equation}
In equation~\ref{eqincnmb}, we assume that there are $K$ uptake rates (where $K = \lvert U \rvert $), and the NMB is calculated for the $i^{th}$ uptake rate $u_i$, assuming that $u_{i-1} = \max\{u_j: j \neq i, u_j < u_i\}$, without loss of generality. We propose that the `critical' uptake rate, which may be denoted by $u^*$ can be determined as the uptake rate at which $NMB(u_i)$ falls below some predetermined threshold $T_{NMB}$. 
A natural choice for $T_{NMB}$ is 0, which is often the threshold used to characterize a given health intervention as being cost-effective or not \citep{prinja2018health}. This is expressed in equation~\ref{eqcritnmb} below.
\begin{equation}
    \label{eqcritnmb}
    u^* = \underset{u_i\in U}{\arg}\min \{NMB(u_i) \vert NMB(u_i) \geq T_{NMB}\} 
\end{equation}
The $NMB(u_i)$ values are provided in Figure \ref{figinmb} below for each uptake rate greater than 10\%. It is clear that there appears to be limited health economic benefit to be gained by increasing $u_i$ beyond 95\%. We acknowledge, however, that assessing the value of expanding (reimbursable) access to treatment to even one additional person requires careful judgement and involves a certain degree of subjectivity. This is reflected in the fact that $T_{NMB}$ in equation~\ref{eqcritnmb} must still be specified by the analyst/policymaker.%$NMB(u_i)$ is a decreasing function of $u_i$ and that it is close to 0 at $u_i = 95\%$. Thus

At the patient level, there is a substantial, statistically significant improvement in all outcomes (health outcomes increase and costs decrease) as uptake rate increases. The rates of change of LYs and  QALYs are convex, peaking at 50\% and decreasing with further increase in uptake rate - once again indicating diminishing returns. The most striking result is the improvement in  QALYs – an improvement of 30.0\% in undiscounted  QALYs and 26.0\% in discounted QALYs, reaffirming that increasing penetration of treatment among those chronically infected leads to more significant gains in quality of life than in life expectancy itself. %Overall, at the patient level as well, there appear to be limited gains in health outcomes - in particular, in  QALYs - beyond a 90\% uptake rate. %Compared to an increase in undiscounted life years of 11.1\% and 10.0\% in discounted life years, it is evident that, as expected, 

The above outcomes clearly suggest that efforts to increase treatment uptake rate, up to a certain threshold, yield substantial epidemiological, health and economic benefits. Our analysis suggests an approach (expressed in equations~\ref{eqincnmb} and~\ref{eqcritnmb}) for determining the critical uptake rate beyond which incremental health economic benefits are minimal. 

\begin{figure}[htp]
	\centering
	\includegraphics[width=12cm]{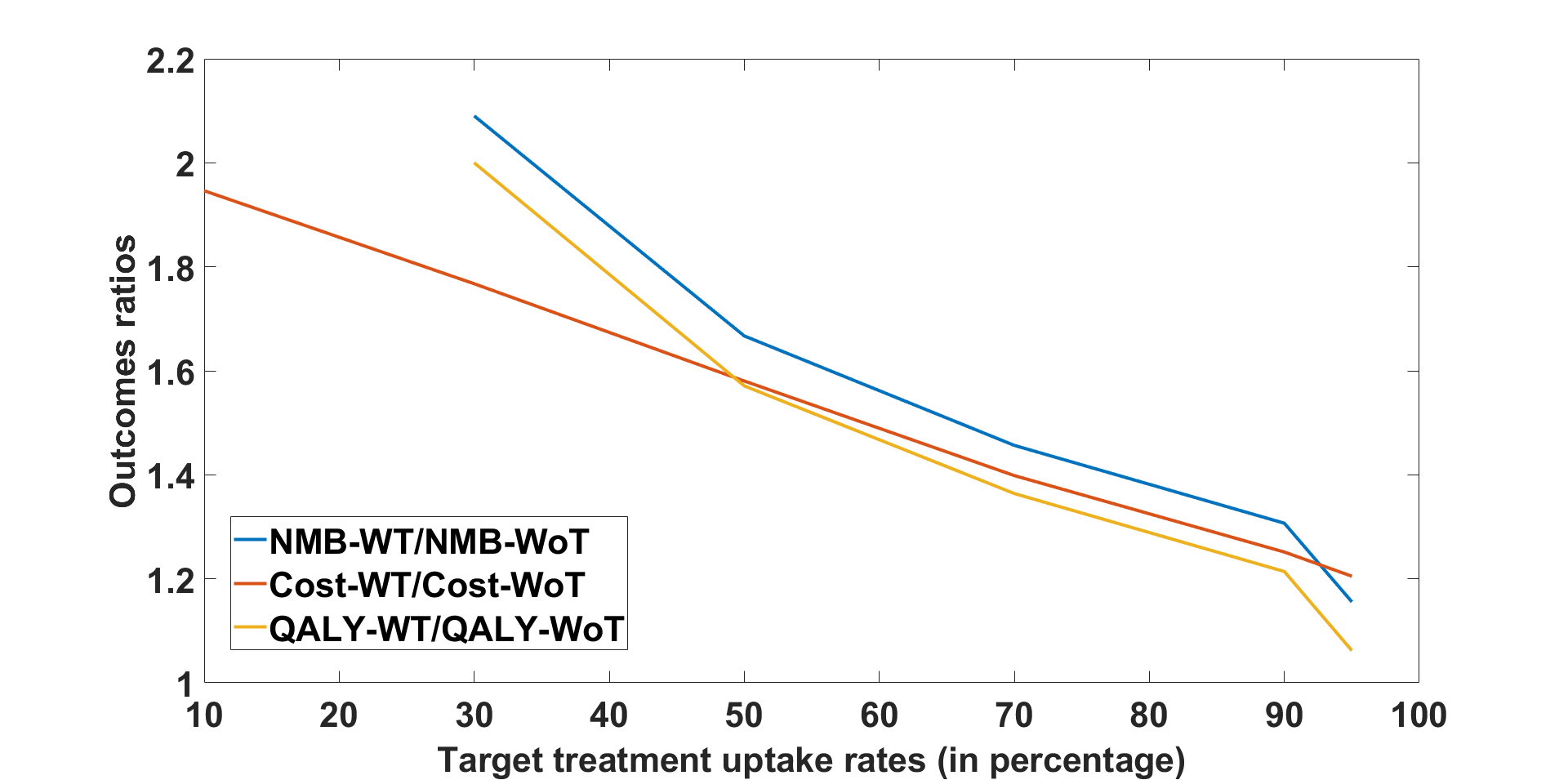}
	\caption{Ratios of outcomes - net monetary benefits (NMBs), costs and the QALY terms within the NMB equation - for the with transmission (WT) and the without transmission (WoT) analyses.}
	\label{fig:ratios}
\end{figure}

We now discuss outcomes that quantify the extent to which incorporating transmissions impacts the cost-effectiveness of HCV treatment. These outcomes were obtained by stopping transmissions during the intervention period of a given simulation run. This involved modifying the values of the transmission-related parameters - for example, the probability of an IDU influencing a non-IDU into becoming an IDU is set to zero. All other simulation settings are retained, and the outcomes from this simulation experiment are depicted in Figure~\ref{fig:ratios}. In Figure~\ref{fig:ratios}, we plot the ratios of (a) the NMBs from the WT to the WoT analyses, (b) the costs from the WoT to the WT analyses (for ease of comparison with the other ratio plots), and (c) the QALY terms from the NMB equation for the WT and the WoT cases. It is clear that in all cases, the ratios converge to 1 as the treatment uptake rate increases. The numerical outcomes corresponding to Figure~\ref{fig:ratios} are provided in Table~\ref{tabwithouttransmission} below.

\begin{table}[htbp]
	\centering
	\caption{Impact of treatment uptake rate - \textit{without} incorporating transmissions in the analysis - on the cost-effectiveness of directly-acting antivirals. Standard deviations in parentheses.}
	\label{tabwithouttransmission}
	\resizebox{\textwidth}{!}{
		\begin{tabular}{|c|cccccc|}
			\hline
			\hline
			Target treatment uptake rate (\%) & 10\%  & 30\%  & 50\%  & 70\%  & 90\% & 95\% \\
			Effective treatment uptake rate (\%) & 9.0 (0.4) & 27.7 (0.7) & 46.7 (0.6) & 65.7 (0.8) & 83.5 (0.2) & 87.5 (0.4) \\
			\hline
			\hline
			\multicolumn{1}{|c}{\textit{Population-level outcomes}} &       &       &       &       &  & \\
			\hline
			Life years & 70.15 (0.04) & 70.17 (0.04) & 70.23 (0.05) & 70.29 (0.04) & 70.30 (0.03) & 70.35 (0.03) \\
			Life years (discounted) & 26.94 (0.01) & 26.96 (0.01) & 26.99 (0.01) & 27.02 (0.01) & 27.04 (0.01) & 27.05 (0.01) \\
			QALYs (years) & 59.53 (0.03) & 59.58 (0.03) & 59.66 (0.04) & 59.74 (0.03) & 59.80 (0.02) & 59.84 (0.02) \\
			QALYs (years, discounted) & 23.63 (0.01) & 23.66 (0.01) & 23.70 (0.01) & 23.74 (0.01) & 23.77 (0.01) & 23.79 (0.01) \\
			Costs (INR) & 16233 (156) & 13813 (173) & 11350 (301) & 8860(160) & 6384 (47) & 5833 (139) \\
			Costs (INR, discounted) & 6646 (72) & 5829 (47) & 4998 (99) & 4156 (38) & 3324 (27) & 3137 (28) \\
			Net monetary benefit & 0     & 11074 & 25581 & 40099 & 51188 & 58213 \\
			\hline
			\multicolumn{1}{|c}{\textit{Patient-level outcomes}} &       &       &       &       &  & \\
			\hline
			Life years & 68.99 (0.45) & 71.03 (0.43) & 73.65 (0.27) & 76.10 (0.56) & 77.86 (0.24) & 78.53 (0.62) \\
			Life years (discounted) & 38.53 (0.24) & 39.81 (0.22) & 41.23 (0.16) & 42.61 (0.24) & 43.70 (0.22) & 44.02 (0.3) \\
			QALYs (years) & 46.00 (0.36) & 49.50 (0.35) & 53.45 (0.13) & 57.29 (0.47) & 60.60 (0.28) & 61.46 (0.63) \\
			QALYs (years, discounted) & 26.99 (0.23) & 28.85 (0.21) & 30.81 (0.12) & 32.75 (0.24) & 34.46 (0.22) & 34.88 (0.32) \\
			Costs (INR) & 814580 (4020) & 692810 (8891) & 569150 (13223) & 444425 (7936) & 320057 (3031) & 292583 (9115) \\
			Costs (years, discounted) & 333493 (1194) & 292370 (1562) & 250647 (3950) & 208454 (1345) & 166623 (320) & 157317 (2316) \\
			\hline
			\hline
		\end{tabular}%
	}%
\end{table}%

%First, we note that discussing epidemiological outcomes in this context is not meaningful, given that transmissions are not considered in generating the outcomes in Figure~\ref{fig:ratios} and Table~\ref{tabwithouttransmission}. Thus, we directly discuss health and economic outcomes. 
We see that, for all uptake rates, the NMBs from the \textit{with transmission} ($WT$) case are significantly larger than those from the \textit{without transmission} ($WoT$) case. In fact, we see that for the 30\% uptake rate case, the NMB for the $WT$ case is more than double the $WoT$ case. The ratio of the NMBs for the $WT$ and $WoT$ cases decreases with increase in uptake rate and approaches one. This is expected, as with an uptake rate of 100\%, no transmissions are possible given that all HCV patients are treated. Thus, the key insight that emerges from this experiment is that incorporating transmissions into assessments of the cost-effectiveness of HCV treatments is vital especially when treatment uptake rates are low (e.g., $<$ 50\%), as is the case in India \citep{cdc24}. %For instance, the ratio is 1.16 for an uptake rate of 95\%. %Interestingly, we also see from Tables~\ref{tabwithouttransmission} and~\ref{tabwithtransmissions} that the rate of increase of the NMBs for the $WT$ case is significantly smaller than that for the $WoT$ case. For example, the increase in NMB when uptake rate increases from 90\% to 95\% for the $WT$ case is just 0.5\% compared to an increase of 13.7\% for the $WoT$ case. This is because a larger improvement in QALYs is observed in the $WoT$ case compared to the $WT$ case as uptake rate increases. This in turn is driven by the fact that incorporating transmissions leads to a smaller rate of increase in health benefits (because they are measured across the entire population, including uninfected agents), especially as uptake rate increases. 

We also observe that medical costs are significantly higher in the $WT$ case - at a 10\% uptake rate, we see that the ratio between the discounted medical costs for the $WT$ case to that of the $WoT$ case is 1.95, and decreases to 1.25 when the uptake rate is 90\%. This increase in medical costs when transmissions are considered is due to the fact that more infections are created and therefore more patients are also treated. The question then arises as to how the NMBs are substantially larger for the $WT$ case when compared to the $WoT$ case. Two factors drive this. First, for a given uptake rate, the health gains (improvement in QALYs) - with respect to the default uptake rate (10\%) - are higher for the $WT$ case than those for the $WOT$ case. Secondly, we recall that the NMBs for each uptake rate associated with the $WT$ and $WoT$ cases are calculated with respect to the outcomes associated with 10\% annual treatment uptake rate for the same case, and not with respect to the other $WT/WoT$ case. Therefore, in addition to the improved health outcomes for the $WT$ case, we also see that the medical cost savings achieved at a population level when treatment uptake rate is increased is substantially larger for the $WT$ case than for the $WoT$ case. %This is reflected in the fact that, for the same 50\% uptake rate, the contribution of the costs term in equation~\ref{eqnmb} for the $WT$ case is approximately 3.06 times that for the $WoT$ case. %For example, at a 50\% uptake rate, the mean discounted QALYs is 23.78 ($SD$ = 0.01) for the $WT$ case, compared to 23.70 ($SD$ = 0.01) for the $WOT$ case. While this appears to be a marginal difference, when multiplied with the willingness-to-pay threshold (3 times the per-capita $GDP$, amounting to $INR$ 341,901) this translates to a substantially large difference between the $WT$ and $WoT$ cases. For instance, the contribution of the health benefits term from equation~\ref{eqnmb} to the NMB for the $WT$ case is approximately 1.6 times that for the $WoT$ case.

%All of the above trends are summarized in Figure~\ref{fig:ratios}. 
%Finally, we observe that the patient-level health (in terms of life years and QALYs) as well as economic outcomes  are higher for the $WT$ case compared to those for the $WoT$ case. This is counter-intuitive, because if only chronically infected patients are considered in the calculation of outcomes at a given uptake rate, it is natural to expect that health and economic outcomes would be similar. This improvement in outcomes for the $WT$ case can be explained as follows. When transmissions are incorporated into the analysis, the majority of those infected are relatively younger, because (a) IDUs are modelled to be younger (between 16 and 32 years) \citep{ambekar2008size} and (b) the younger skew of the Indian population - the median age is less than 30 years. Therefore, in the $WT$ case, the cohort of chronically infected patients consists of a larger proportion of younger agents. When treatment is provided to those infected, these younger agents are treated and (in most cases) cured, and hence live slightly longer and with better quality of life. These gains diminish as treatment uptake rate increases, as nearly all infected are treated, and hence this drives the decrease in difference between the NMBs for the $WT$ and $WoT$ case as uptake rate increases.

We conclude the above discussion by reiterating that when treatment uptake rates are low, accounting for HCV transmissions in any assessment of the health and economic benefits of HCV treatment is vital.%, depending on the IDU and general population demographics

\subsection{Research Objective 2 Outcomes: Impact of Timing and Frequency of Treatment}
\label{ro2res}
In this section, we discuss the outcomes of the computational experiments conducted to quantify the impact of varying the timing and frequency of treatment of a slow-moving disease such as HCV - i.e., outcomes associated with \textit{$RO2$}. 

We consider four alternatives to the default $annual-treatment$ model, which represents the status quo treatment model, where awareness regarding HCV increases with time and more patients seek treatment. These alternatives to the \textit{annual-treatment model} function as \textit{camp models}, wherein population-wide treatment camps (including facilities for disease screening) are held once every few years. We implemented four types of camp models:
\begin{enumerate}
    \item Treat only at the beginning (\textit{$T_0$-treatment} model). Here, a certain proportion of all infected agents living at the beginning of the intervention period are treated at that time point. Agents that get infected during the intervention/surveillance period are not treated.
    \item Treat only at the end (\textit{end-treatment} model). Here, a certain X\% of all infected agents living at the end of the intervention/surveillance period are treated at that time point. 
    \item Treat twice (\textit{twice-treatment} model). Here, X\% of all infected agents living at the end of the fifth year and the tenth (final) year of the intervention/surveillance period are treated.
    \item Treat thrice (\textit{thrice-treatment} model). Here, X\% of all infected agents living at the end of the third, sixth year and the tenth (final) year of the intervention/surveillance period are treated.
\end{enumerate}

\begin{figure}
    \centering
\begin{subfigure}{0.4\textwidth}
    \includegraphics[width=\textwidth]{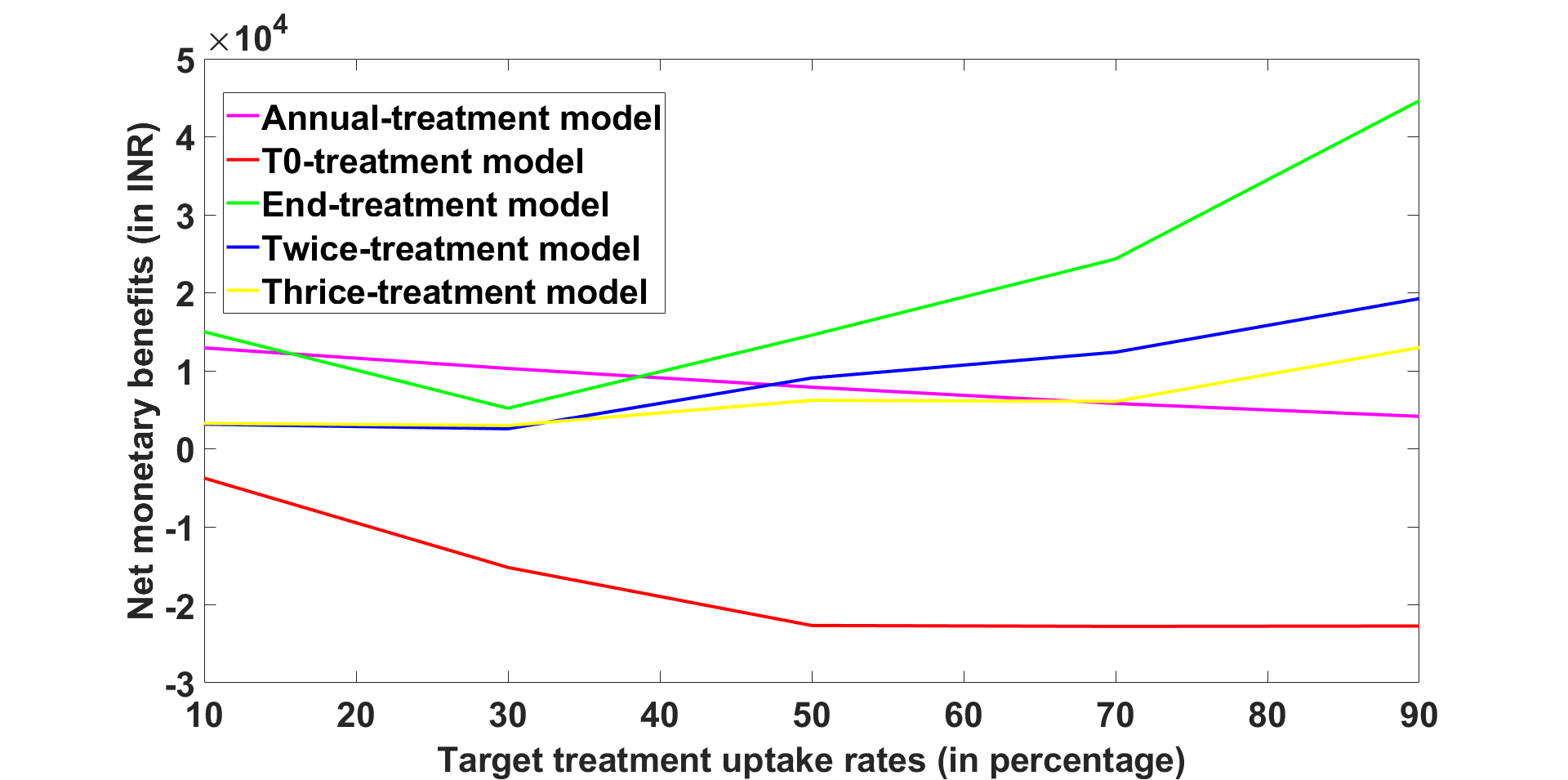}
    \caption{NMBs for alternate treatment models calculated against the \textit{annual-treatment model}.}
    \label{fig:ro2nmb}
\end{subfigure}
\hfill
\begin{subfigure}{0.4\textwidth}
    \includegraphics[width=\textwidth]{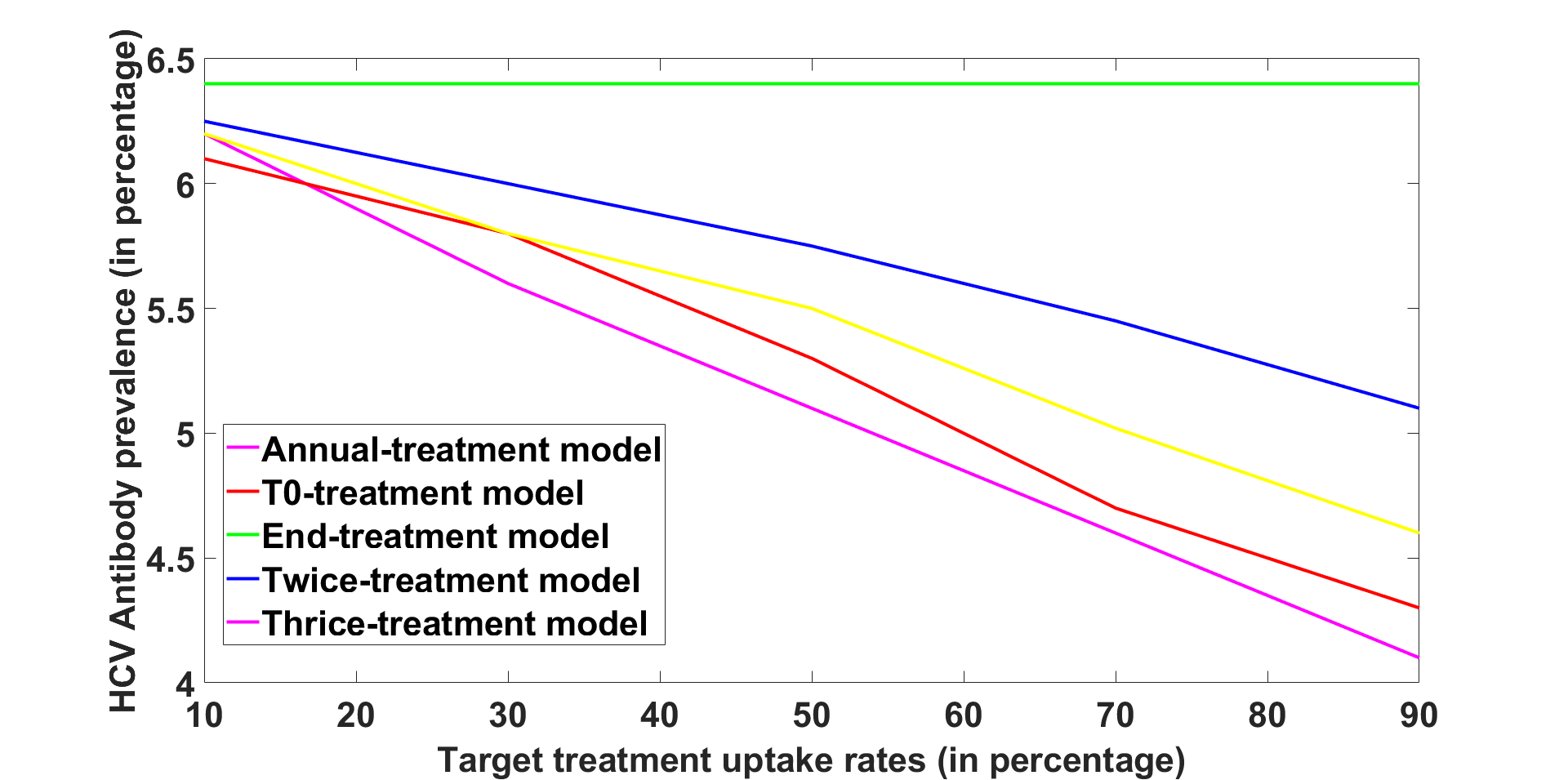}
    \caption{HCV antibody prevalence.}
    \label{fig:ro2ab}
\end{subfigure}
\hfill
\begin{subfigure}{0.4\textwidth}
    \includegraphics[width=\textwidth]{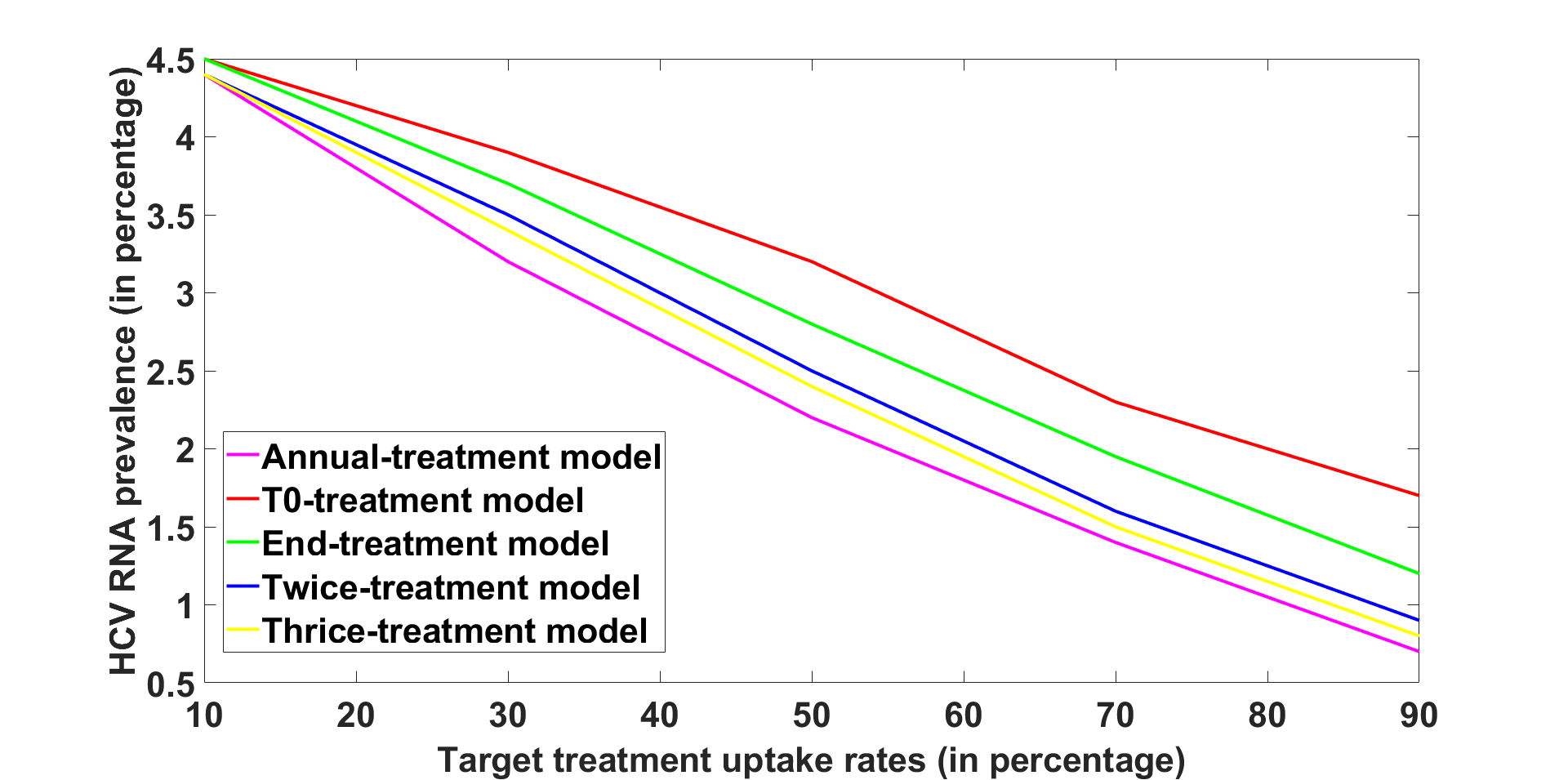}
    \caption{HCV RNA prevalence.}
    \label{fig:ro2rna}
\end{subfigure}
\hfill
\begin{subfigure}{0.4\textwidth}
    \includegraphics[width=\textwidth]{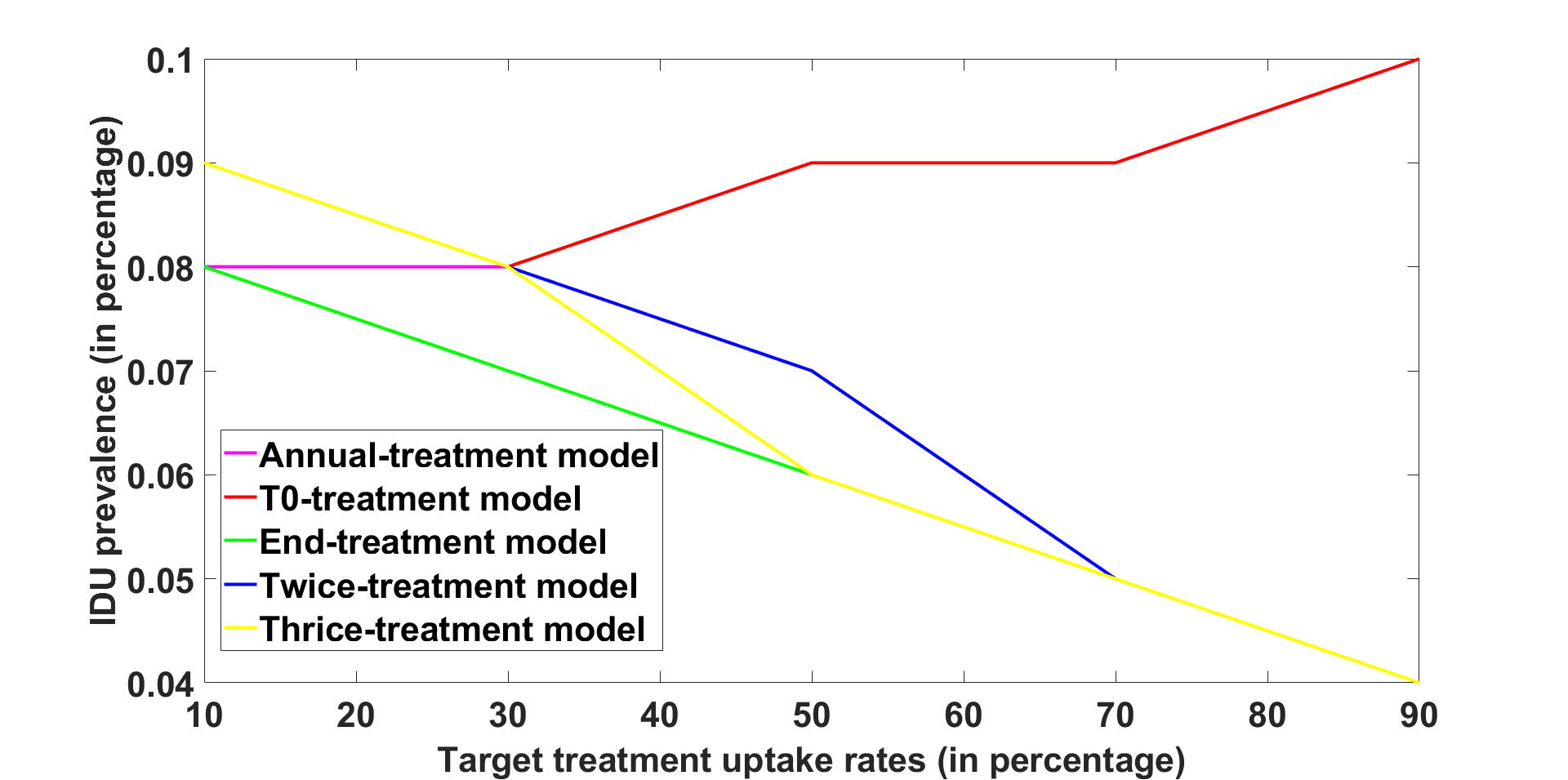}
    \caption{IDU prevalence.}
    \label{fig:ro2idu}
\end{subfigure}    
\caption{Epidemiological outcomes and NMBs associated with multiple treatment models as a function of treatment uptake rate}
\label{fig:leffest}
\end{figure}

Epidemiological outcomes and NMBs associated with the above treatment models are provided in Figures \ref{fig:ro2nmb} through \ref{fig:ro2idu}, and supporting patient-level outcomes are provided in Table \ref{tabobj} below. In Figure~\ref{fig:ro2nmb}, we plot the NMBs of each alternative treatment model calculated against the default \textit{annual-treatment} model. %As an example, for the \textit{end-treatment} model in Figure~\ref{fig:ro2nmb}, each NMB plotted - i.e., for a given uptake rate - is calculated by applying Equation~\ref{eqnmb} to the discounted costs and  QALYs for the \textit{end-treatment} model from Table \ref{tabobj2second}, when compared against the corresponding discounted costs and  QALYs (for the same uptake rate) for the \textit{annual-treatment} model. This provides us with a means to summarize the health and economic outcomes associated with each treatment model in comparison with the default \textit{annual-treatment} model.

\begin{table}[htbp]
	\centering
	\caption{Patient-level health and economic outcomes of treatment models varying in timing and frequency of treatment.}
	\resizebox{\textwidth}{!}{
		\begin{tabular}{|c|ccccc|}
			\hline
			\hline
			Target treatment uptake rate (\%)    & 10\%  & 30\%  & 50\%  & 70\%  & 90\% \\
			\hline
			\hline
			\multicolumn{6}{|c|}{\textit{Annual-treatment} model} \\
			\hline
			Effective uptake rate (\%) & 9.47 (0.23) & 28.98 (0.64) & 47.72 (0.16) & 67.05 (0.84) & 84.89 (0.19) \\
			Life years & 71.04 (0.46) & 72.87 (0.43) & 75.10 (0.17) & 77.10 (0.40) & 78.50 (0.32) \\
			Life years (discounted) & 39.10 (0.21) & 39.93 (0.19) & 41.07 (0.08) & 42.20 (0.21) & 43.12 (0.33) \\
			QALYs & 48.15 (0.35) & 51.58 (0.41) & 55.37 (0.15) & 58.90 (0.27) & 61.76 (0.40) \\
			QALYs (discounted) & 27.68 (0.19) & 29.30 (0.19) & 31.15 (0.06) & 32.96 (0.17) & 34.38 (0.32) \\
			Costs & 795230 (4699) & 676428 (2696) & 543451 (10942) & 420211 (15172) & 303418 (10759) \\
			Costs (discounted) & 326869 (1227) & 282326 (1172) & 235821 (3261) & 192304 (4052) & 152358 (3167) \\
			\hline
			\hline
			\multicolumn{6}{|c|}{\textit{$T_0$-treatment} model} \\
			\hline
			Effective uptake rate (\%) & 4.57 (0.23) & 14.82 (0.79) & 26.81 (0.68) & 42.02 (1.07) & 57.92 (0.24) \\
			Life years & 70.25 (0.37) & 71.44 (0.25) & 72.69 (0.24) & 74.36 (0.70) & 75.66 (0.50) \\
			Life years (discounted) & 38.89 (0.18) & 39.57 (0.16) & 40.34 (0.15) & 41.27 (0.28) & 42.15 (0.28) \\
			QALYs (years) & 46.97 (0.26) & 48.92 (0.27) & 51.02 (0.20) & 53.77 (0.54) & 56.42 (0.41) \\
			QALYs (years, discounted) & 27.22 (0.14) & 28.23 (0.16) & 29.34 (0.15) & 30.74 (0.25) & 32.19 (0.24) \\
			Costs (INR) & 827360 (6353) & 764390 (6566) & 685218 (5449) & 598367 (13420) & 492137 (8066) \\
			Costs (INR, discounted) & 340248 (1707) & 318469 (2045) & 290972 (798) & 260994 (3624) & 223742 (2121) \\
			\hline
			\hline
			\multicolumn{6}{|c|}{\textit{End-treatment} model} \\
			\hline
			Effective uptake rate (\%) & 8.99 (0.19) & 27.26 (0.86) & 45.23 (0.83) & 63.21 (0.88) & 80.26 (0.65) \\
			Life years & 71.07 (0.48) & 73.32 (0.47) & 75.52 (0.33) & 77.74 (0.34) & 80.01 (0.17) \\
			Life years (discounted) & 39.29 (0.21) & 40.63 (0.23) & 41.97 (0.19) & 43.29 (0.21) & 44.58 (0.16) \\
			QALYs (years) & 48.03 (0.32) & 51.55 (0.39) & 54.99 (0.28) & 58.49 (0.29) & 61.93 (0.14) \\
			QALYs (years, discounted) & 27.74 (0.15) & 29.56 (0.20) & 31.39 (0.16) & 33.22 (0.18) & 34.96 (0.13) \\
			Costs (INR) & 802047 (7271) & 685585 (1856) & 575212 (4786) & 459706 (6202) & 354235 (4084) \\
			Costs (INR, discounted) & 331108 (2237) & 292039 (541) & 254975 (1193) & 215820 (1256) & 181015 (842) \\
			\hline
			\hline
			\multicolumn{6}{|c|}{\textit{Twice-treatment} model} \\
			\hline
			Effective uptake rate (\%) & 9.1 (0.2) & 28 (0.6) & 47.2 (0.2) & 66.05 (0.25) & 83.4 (0.3) \\
			Life years & 70.95 (0.41) & 73.16 (0.36) & 75.41 (0.21) & 77.52 (0.23) & 79.24 (0.42) \\
			Life years (discounted) & 39.16 (0.16) & 40.31 (0.17) & 41.52 (0.10) & 42.80 (0.17) & 43.92 (0.20) \\
			QALYs (years) & 47.98 (0.28) & 51.62 (0.23) & 55.37 (0.06) & 58.90 (0.19) & 61.92 (0.37) \\
			QALYs (years, discounted) & 27.67 (0.12) & 29.46 (0.12) & 31.33 (0.05) & 33.19 (0.14) & 34.81 (0.18) \\
			Costs (INR) & 801867 (9231) & 681207 (8684) & 553402 (11234) & 431301 (4697) & 318328 (5344) \\
			Costs (INR, discounted) & 329832 (3262) & 287584 (1936) & 243102 (2396) & 200589 (251) & 162769 (2645) \\
			\hline
			\hline
			\multicolumn{6}{|c|}{\textit{Thrice-treatment} model} \\
			\hline
			Effective uptake rate (\%) & 9.6 (0.4) & 28.6 (0.4) & 47.8 (0.3) & 66.3 (0.3) & 84.3 (0.2) \\
			Life years & 71.20 (0.35) & 73.13 (0.44) & 75.57 (0.23) & 77.70 (0.34) & 79.30 (0.36) \\
			Life years (discounted) & 39.20 (0.20) & 40.21 (0.22) & 41.45 (0.17) & 42.62 (0.18) & 43.69 (0.17) \\
			QALYs (years) & 48.21 (0.25) & 51.74 (0.35) & 55.63 (0.14) & 59.13 (0.20) & 62.22 (0.32) \\
			QALYs (years, discounted) & 27.74 (0.16) & 29.47 (0.19) & 31.36 (0.14) & 33.14 (0.13) & 34.80 (0.16) \\
			Costs (INR) & 800832 (1189) & 673218 (1728) & 547292 (4250) & 433358 (11311) & 311842 (4572) \\
			Costs (INR, discounted) & 328994 (1158) & 284055 (2129) & 239060 (1590) & 199271 (3114) & 158583 (1117) \\
			\hline
			\hline
		\end{tabular}%
	}
	\label{tabobj}%
\end{table}%

% \begin{figure}[htbp]
% 	\centering
% 	\subfloat[NMBs for alternate treatment models calculated against the \textit{annual-treatment model}.]{%
% 		\resizebox*{8cm}{!}{\includegraphics{NMBs objective 2 plot.png}} \label{fig:ro2nmb}}\hspace{5pt}
% 	\subfloat[HCV antibody prevalence.]{%
% 		\resizebox*{8cm}{!}{\includegraphics{Antibody objective 2 plot.png}}\label{fig:ro2ab}}\hspace{5pt}
% 	\subfloat[HCV RNA prevalence.]{%
% 		\resizebox*{8cm}{!}{\includegraphics{HCV RNA objective 2 plot.png}}\label{fig:ro2rna}}\hspace{5pt}
% 	\subfloat[IDU prevalence.]{%
% 		\resizebox*{8cm}{!}{\includegraphics{IDU objective 2 plot.png}}\label{fig:ro2idu}}\hspace{5pt}
% 	\caption{Epidemiological outcomes and NMBs associated with multiple treatment models as a function of treatment uptake rate.}
% 	\label{fig:leffest}
% \end{figure}

It is evident from Figure \ref{fig:ro2nmb} that the \textit{$T_0$-treatment} model is the least cost-effective model. This implies that instituting screening and treatment programmes - especially when resources are limited - at the beginning of a surveillance period is not optimal in terms of NMBs. In fact, for a slow-moving disease such as HCV, such a model is clearly inferior in comparison to the \textit{end-treatment} model in terms of economic outcomes. From Table~\ref{tabobj}, it can be seen that the default \textit{annual-treatment} model yields the least costs; however, it also yields the lowest per-patient  QALYs as well compared to the other models (except the \textit{$T_0$-treatment} model).% The difference in  QALYs is marginal - for example, the difference between default model and the best-performing model in terms of  QALYs is only 0.8\%. When combined with the willingness-to-pay threshold, however, this marginal difference contributes to making this model the most cost-effective when NMBs are considered.

From Figure~\ref{fig:ro2nmb}, it is evident that if resource constraints exist such that only one screening and treatment camp can be held during a surveillance period for a disease with a progression profile similar to HCV, treating at the end of the surveillance period is more beneficial. However, the speed of spread of the disease in question is also important here. In our modeled scenario, HCV is not only slow-progressing but also relatively slow in its spread: for example, under the \textit{end-treatment} model, without any treatment for a period of 10 years, we see that HCV antibody prevalence increases from 3.6\% to only 6.4\%, which corresponds to an approximately 6\% annual increase in the number of infections. Alternatively, for a disease that spreads or progresses faster, an intervention such as the \textit{end-treatment} model may not be sufficiently (cost-) effective. Thus the optimal timing of screening and treatment for a disease will depend on its speed of progression as well as speed of spread.%Importantly, this is the case for an Indian province (Punjab) where HCV prevalence and spread is much higher than the national average.

With regard to the frequency of treatment, it is immediately apparent that all treatment models other than the \textit{$T_0$-treatment} model yield positive NMBs in comparison to the default model. This implies that the treatment camp model - whether it is conducted twice, thrice or once at the end of a 10-year period - is superior (in NMB terms) to conducting it every year, especially if the diagnosis and treatment rate for HCV is low without screening or awareness programmes. Among these camp models, the \textit{end-treatment} model appears to be the most effective, with its NMBs outstripping the other models especially as uptake rates increase (e.g., beyond 30\%). %At low uptake rates (below 20\%), however, all three models appear to yield marginal benefits over the default model. 

We evaluated two other camp models to further investigate and verify the above findings on the impact of timing of treatment on the cost-effectiveness of DAAs. The first treatment model was an alternate version of the $twice-treatment$ model, wherein treatment was implemented at the beginning of the intervention period (year 0) and at the end of the 5\textsuperscript{th} year, as opposed to at the end of the 5\textsuperscript{th} and 10\textsuperscript{th} years. The second treatment model was, similarly, an alternate version of the \textit{thrice-treatment} model, with treatment at the end of the 0\textsuperscript{th}, 3\textsuperscript{rd} and 6\textsuperscript{th} years, as opposed to at the end of the 3\textsuperscript{rd}, 6\textsuperscript{th} and 10\textsuperscript{th} years. As done before, we computed the NMBs of these treatment models against that of the \textit{annual-treatment} model for each value of uptake rate up to 90\%, and found that both treatment models yielded negative values of the NMB at every uptake rate when compared to the \textit{annual-treatment} model. 

Therefore, the following key insight emerges from $RO2$: as uptake rates increase, within a surveillance/intervention period that is significantly shorter (e.g., $<$ 50\%) than the life expectancy of patients with the disease, it is evident that screening and treating less frequently appears beneficial. In particular, the importance of conducting a treatment camp at the end of the surveillance period is clearly emphasized by the above results. 

We must emphasize here that while screening and treating less frequently (i.e., the \textit{end-treatment} or the \textit{twice-treatment} models) appears more cost-effective, doing so does yield significantly larger numbers of infections (see Figures~\ref{fig:ro2ab} - \ref{fig:ro2idu}). Thus, if the goal is to eradicate HCV, then the default model remains the most effective with as high an uptake rate as possible.%For example, the antibody prevalence with the \textit{twice-treatment} model is nearly 10\% higher when compared to the \textit{thrice-treatment} model, and nearly 20\% higher when compared to the default model. 

Further, there are other effects of adopting treatment models that lead to larger numbers of HCV infections that may be worth considering prior to decision-making. For example, HCV infections are often accompanied by HIV co-infections, especially among injecting drug users. Similarly, given that the majority of infected IDUs are younger in age than the average Indian person, larger numbers of HCV infections mean more young people who are ineligible to donate blood or organs. If such `second order' effects are unlikely to occur, then the \textit{twice-treatment} or the \textit{end-treatment} models appear to be the most cost-effective options - this is particularly the case if the diagnosis and treatment rate without camps is low. 

\section{Discussion}
\label{sec:disc}
The first question that we attempt to answer in this study, via \textit{RO1}, is: to what extent is the cost-effectiveness of DAAs underestimated if transmissions are not considered in a model designed for this purpose? The extent of penetration of treatment among those infected plays a key role in its epidemiological effectiveness and therefore its cost-effectiveness, and hence we answer this question by quantifying the extent of underestimation as a function of treatment uptake rate. We find that the extent of underestimation is considerable - exceeding 100\% - at lower uptake rates than at higher uptake rates. Treatment uptake rates are low in India \citep{who24}, with recent studies estimating it to be 21\%, and hence our findings provide further evidence to support continued subsidization of HCV treatment with DAAs, especially in regions with prevalence comparable to that in the Punjab province (approximately 3\%). These findings are likely applicable to other developing countries where prevalences are at a similar or higher level: for example, Pakistan, certain African countries and Uzbekistan, among others \cite{lim2020effects,cdc24}. From a modeling standpoint, this indicates that incorporating transmissions analyses situated in regions with high uptake may be less important than in regions with low uptake. Once again, this implies that cost-effectiveness analyses studies conducted for DAAs in regions where uptake rates are low need to incorporate transmissions to accurately capture their health and economic value. 

As part of \textit{RO1}, we also provide a method to determine the extent to which the uptake of treatment must be increased by computing the \textit{incremental NMB} associated with a given uptake rate. As mentioned before, while the question of determining the extent to which treatment uptake must be increased is controversial, our approach facilitates making this decision in resource-constrained health systems. This approach also aids decision-making around which population subgroups to prioritize for provision of subsidized treatment - for example, relatively less privileged groups may be prioritized. Once again, this approach may be more broadly applicable - for example, to the case of expanding treatment access for HIV/AIDS or tuberculosis.

The second set of questions that we attempt to answer through this study, as part of \textit{RO2}, is around the timing and frequency of treatment within a disease surveillance period. The key insight that emerges from this analysis is that the speed at which the disease progresses in infected individuals as well as the speed with it spreads are vital to determining the optimal timing and frequency of organizing treatment drives in terms of overall economic benefit. For instance, for a slow-moving disease such as HCV, our analysis indicates that at least one treatment drive must be held at the end of the surveillance period to achieve maximum health and economic value. However, from an epidemiological standpoint - i.e., in terms of minimizing the number of infections, or eradication of the disease - treating only at the end of the surveillance period may not be effective.

A limitation of our analysis around \textit{ROs} 1 and 2 involves the lack of longitudinal data to validate the trends in HCV epidemiology generated by the ABS. However, our medical collaborator with substantial clinical and public health experience in the treatment and surveillance of HCV in the state of Punjab provided face validation to the longitudinal outcomes generated by the model. Our analysis also did not incorporate indirect medical costs - such as those of the productivity losses of patients and their caregivers - due to limited availability of data. Addressing this limitation, and the conduct of a cost-effectiveness analysis of screening programs, provide avenues of future research. Another key avenue of future work, motivated by the findings from \textit{RO2}, involves further investigation of the relationship between: (a) speed of infectious disease progression; (b) its spread, and (c) optimal frequency and timing of screening. 

Overall, our study provides guidance for both policymakers and modelers, in terms of actionable insights regarding HCV, and more generally for slow-moving and relatively slow-spreading (in comparison to airborne pathogens) infectious diseases, such as HIV, where similar problems have been considered before in a deterministic setting, as in \cite{deo2015planning}. These include insights regarding screening and treatment program management for policymakers, and insights regarding when to incorporate transmissions and more efficient methods for outcomes estimation for modelers.

\bibliographystyle{model5-names}
 \bibliography{cas-refs}

 \appendix

 \section{Hepatitis C Virus Infection Progression Module}
\label{appdisprog}
\begin{figure}[t]
	\begin{center}
		\includegraphics[height=3.5cm]{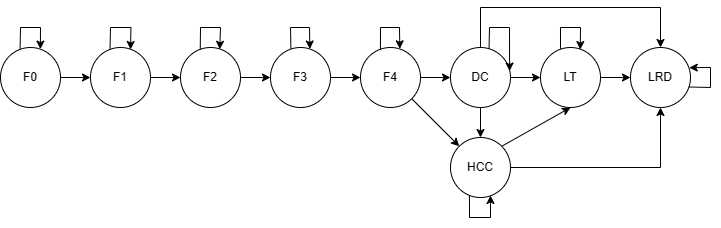}
		\caption{The discrete-time Markov chain used to model the natural history of chronic HCV in an infected agent.} 
		\label{figdtmc}
	\end{center}
\end{figure}

Approximately 26\% of those newly infected with HCV clear the infection within six months of being infected \citep{micallef2006spontaneous}. This is the acute state of HCV. Our clinical collaborator confirmed that the health and economic impacts of HCV were primarily associated with the chronic stages, and this is also supported in the literature \citep{chugh2019real}. The chronic states are divided into two groups - (i) early chronic states, and (ii) critical and advanced states. The early states include - in chronological order from first to last - the F0, F1, F2, F3 and F4 states wherein the liver is not damaged beyond cure. F4 is the first state of cirrhosis (compensated cirrhosis), and is followed by the first state in the second group - DC or decompensated cirrhosis. Following the DC state is HCC or hepatocellular carcinoma, which incurs substantial one-time as well as recurring management costs. Many patients in the DC and the HCC states have to undergo a liver transplant, and hence we incorporate this state as well in the DTMC. Finally, patients of states DC, HCC or even those who have had liver transplants can die of liver-related causes, and hence we incorporate an absorbing state for liver-related deaths (the LRD state). This DTMC is depicted in Figure~\ref{figdtmc}. % and the mechanism of clearing the HCV infection within six months of being infected without treatment is called \textit{spontaneous clearance}% Even though DC is an advanced cirrhosis state, DAAs are able to cure patients in this state with high effectiveness \citep{chugh2019real,chugh2021cost}. 

Infected agents can exit the fibrosis states (F0 - F3), the F4 and the DC states if they are cured - i.e., achieve sustained virologic response or SVR - via treatment with DAAs. Patients who achieve SVR from the cirrhotic states (DC or F4) may still experience transition to the DC or HCC states, albeit with a significantly reduced probability than if they still were infected, due to the damage their liver has already sustained.

The NH-DTMC is implemented as a Monte Carlo simulation for each individual infected agent to capture heterogeneity of outcomes possible with HCV. In other words, the progression of HCV for each infected agent can be considered as an $iid$ replication of this simulation, conditional on the age of infection (or state of infection, if the agent is part of the initial cohort of infected agents). %The transition probabilities for the NH-DTMC, along with the probabilities for transitioning to the DC and HCC states after achieving SVR from the cirrhotic states are provided in Table 2 in section 2 of the e-component. %The algorithm capturing the execution of the simulation replication for each infected agent is also provided as Algorithm 1 in section 2 of the e-component.

The annual transition probabilities for the NH-DTMC (see Section \ref{sec3}) are given in Table \ref{tableprog}. The annual transition probability from DC to LRD (0.182) was higher in the first year than in the subsequent years (0.112). Similarly, the annual transition probability from LT to LRD was higher in the first year (0.116) than in the subsequent years (0.044). All the other transition probabilities remained constant over the sojourn of an agent in a particular state. We also incorporate the annual transition probabilities from the state SVR2 to states DC and HCC as discussed above in this section.

\begin{table}[htbp]
    \centering
     \caption{State transitions and transition probability estimates, for the natural history discrete-time Markov chain, obtained from \cite{chhatwal2015cost}.} \label{tableprog}
     \resizebox{0.7\textwidth}{!}{
    \begin{tabular}{|c|c|}
    \hline
    \textbf{Transition} & \textbf{Probability} \\ [0.5ex]
    \hline
    Acute - Healthy & 0.26 (Six months) \\
    \hline
    F0-F1; F1-F2; F2-F3; F3-F4 & 0.117; 0.085; 0.120; 0.116 \\
    \hline
    F4-DC; F4-HCC & 0.039; 0.014 \\
    \hline
    DC-HCC; DC-LT; DC-LRD & 0.068; 0.023; 0.182/0.112 (first year/after first year) \\
    \hline
    HCC-LT; HCC-LRD & 0.040; 0.427 \\
    \hline
    LT-LRD & 0.116/0.044 (first year/after first year) \\
    \hline
    SVR2-DC; SVR2-HCC & 0.008; 0.005 \\
    \hline
    \end{tabular}
    }
\end{table}

The implementation of the NH-DTMC and the non-DRM models as a Monte Carlo simulation replication for each infected agent is captured in Algorithm~\ref{algupdt} below.

\begin{algorithm}
	\caption{Infection status and health state update (execution of the NH-DTMC and non-DRD models) subroutine.}
	\label{algupdt}
	\begin{algorithmic}[1]
		\State \textbf{Begin Initialization} \Comment{Initialization phase}
		\State Initialize with $K \times K$ transition probability matrix $P$ for the NH-DTMC.
		\State Construct $cdf$ for random variables $X_k$, where $X_k$ represents the disease state in the next time point, for each $k \in \{1, 2, \dots, K\}$.
		
		\For{each $k \in \{1, 2, \dots, K\}$}
		\State Identify the row of $P$ corresponding to $k$.
		\State Identify non-zero entries of $P$.
		\State Construct $cdf$ $F_k(x)$ corresponding to state $k$.
		\State \hspace{10pt} Example: for disease state $k$, $k$-th row of $P$ is identified.
		\State \hspace{10pt} Let non-zero entries be $p_{k1}$, $p_{k2}$, ..., $p_{kK}$ ($k \leq K$).
		\State \hspace{10pt} Step points for $cdf$ $F_{k}(x)$: $p_{k1}$, $p_{k1} + p_{k2}$, $p_{k1} + p_{k2} + p_{k3}$ and so on.
		\EndFor
		
		\State \textbf{End Initialization} \Comment{End of initialization phase}
		
		\State Current infection status of agent $Inf_t := 1$.
		\State Current disease state of agent $k_t := k$.
		
		\While{agent is alive AND $Inf_t = 1$}
		\State Execute non-DRM model: if agent is alive, proceed; else, BREAK.
		\State Is patient cured? If yes, $Inf_t = 0$, BREAK; else $Inf_t = 1$, proceed.
		\State Identify $cdf$ $F_k(x)$ corresponding to $k$.
		\State Sample $u$ = $U(0,1)$.
		\State Set new state as $k_t = F_k^{-1}(u)$.
		\EndWhile
	\end{algorithmic}
\end{algorithm}

 \section{Treatment Module of the Agent-based Simulation
Model: Additional Details}
 \label{aptrt}

In Table \ref{tabanntreat}, we provide an example of the implementation of the \textit{annual-treatment} algorithm (described in Section \ref{treatmod}). Untreated infections in every time period are treated according to its corresponding coverage rate. For example, of 20 infections caused at time point 1, 14 are assigned to treatment. At $j = 1$, 52\% of these 14 infections are treated. Of the remaining infections among these, 64\% are treated at time point 2. Bold-faced numbers indicate that the total number of new infections \textit{and} treatment assignments in each time period matches the respective number of treatment assignments. 
 
 \begin{table}[htbp]
  \centering
  \caption{Example illustrating the implementation of the \textit{annual-treatment} algorithm. $u = 0.7, \alpha = 0.4, \text{ and } n = 5$.}
  \resizebox{\textwidth}{!}{
    \begin{tabular}{|c|c|c|c|c|c|c|c|c|c|}
    \hline
    Time period & Number of new infections & Number of new infections & Coverage fraction  & \multicolumn{6}{c|}{Number of infections treated } \\
          &       & assigned to treatment & for & \multicolumn{6}{c|}{among those caused at time point} \\
\cline{5-10}          &       &    & treatment-assigned patients   & \multicolumn{1}{c}{0} & \multicolumn{1}{c}{1} & \multicolumn{1}{c}{2} & \multicolumn{1}{c}{3} & \multicolumn{1}{c}{4} & 5 \\
    \hline
    0     & 100   & \textbf{70} & 0.40 & 28.00    & -     & -     & -     & -     & - \\
    1     & 20    & \textbf{14} & 0.52 & 21.84 & 7.28  & -     & -     & -     & - \\
    2     & 16    & \textbf{11.2} & 0.64 & 12.90 & 4.30 & 7.17 & -     & -     & - \\
    3     & 12    & \textbf{8.4} & 0.76 & 5.52 & 1.84 & 3.06 & 6.384 & -     & - \\
    4     & 8     & \textbf{5.6} & 0.88 & 1.53 & 0.51 & 0.85 & 1.774 & 4.93 & - \\
    5     & 4     & \textbf{2.8} & 1.00 & 0.21 & 0.07 & 0.12 & 0.242 & 0.67 & 2.80 \\
    \hline
    Total & 160   & 112   & \textbf{\--\--} & \textbf{70} & \textbf{14} & \textbf{11.2} & \textbf{8.4} & \textbf{5.6} & \textbf{2.8} \\
    \hline
    \end{tabular}%
    }
  \label{tabanntreat}%
\end{table}%

We now provide a straightforward mathematical argument to show that Algorithm~\ref{alganntrt} yields the required treatment uptake rate under deterministic conditions. For this, it is sufficient to show that among infections caused at any time point $j_1$, 0 $\leq$ $j_1$ $\leq$ $n$, the number of these infections undergoing treatment will be equal to the number of these infections assigned to treatment. In other words, we need to show:
\begin{equation}
   \sum_{j=j_1}^{n} N^{Tr}_{j_1,j} = N^a_{j_1}, \hspace{2pt} j_1= \hspace{2pt} 0, \hspace{2pt} 1, \hspace{2pt}..., \hspace{2pt} n
\end{equation}

We first note that the coverage fraction for treatment-assigned patients for time period $j_1$ is [$\alpha + \{(1-\alpha) \times \frac{j_1}{n}\}]$, the number of infections created and assigned to treatment that are also treated at $j_1$ is given by:
\begin{equation}
\label{proofata1}
    N^{Tr}_{j_1,j_1} = [\alpha + \{(1-\alpha) \times \frac{j_1}{n}\}] \times N^a_{j_1}
\end{equation}
Consider an agent that acquires an infection at $j_1$. Assume that the event that it does not get treated at $j_2$, where $0 \leq j_1 < j_2 \leq n$, is independent of the event that it has been untreated till time point $(j_2 - 1)$. The probability that any treatment-assigned infection created at time point $j_1$ gets treated at a time point $j$, where $0 \leq j_1 < j < n$ is the coverage fraction $[\alpha + \{(1-\alpha) \times \frac{j_1}{n}\}]$. Hence, the probability that any treatment-assigned infection (created at time point $j_1$) does not get treated at time point $j$, where $0 \leq j_1 < j \leq n$ is given by the expression:
\begin{equation*}
    1 - [\alpha + \{(1-\alpha) \times \frac{j}{n}\}] = [(1 - \alpha) - \{(1-\alpha) \times \frac{j}{n}\}]
\end{equation*}
Thus, the probability of any treatment-assigned infection (created at time point $j_1$) not being treated till time point $(j_2 - 1)$, where $0 \leq j_1 < (j_2 - 1) \leq n$, is given by the expression:
\begin{equation*}
    \prod_{j = j_1}^{j_2 - 1} [(1 - \alpha) - \{(1-\alpha) \times \frac{j}{n}\}]
\end{equation*}
Moreover, we assume that the event of the agent being treated at time point $j_2$ is independent of the event that it has not been treated until the time point $(j_2 - 1)$. Then the probability of the agent not being treated until time point $(j_2 - 1)$ and then being treated at $j_2$ is given by:
\begin{equation}
    \prod_{j = j_1}^{(j_2 - 1)} [(1 - \alpha) - \{(1-\alpha) \times \frac{j}{n}\}] \times [\alpha + \{(1-\alpha) \times \frac{j_2}{n}\}]
\end{equation}
Thus, the number of infections caused at time point $j_1$ and treated at time point $j_2$ is given by:
\begin{equation*}
	\begin{aligned}
    N^{Tr}_{j_1,j_2} &= \prod_{j = j_1}^{j_2 - 1} [(1 - \alpha) - \{(1-\alpha) \times \frac{j}{n}\}] \times [\alpha + \{(1-\alpha) \times \frac{j_2}{n}\}] \times N^a_{j_1}, \hspace{2pt} 0 \hspace{2pt} \\
    & \leq \hspace{2pt} j_1 \hspace{2pt} < \hspace{2pt} (j_2 - 1) \hspace{2pt} \leq \hspace{2pt} n \\
    \end{aligned}
\end{equation*}
If $j_2$ = $n$, then the above equation becomes:
\begin{equation}
\label{proofata}
    N^{Tr}_{j_1,n} = \prod_{j = j_1}^{n - 1} [(1 - \alpha) - \{(1-\alpha) \times \frac{j}{n}\}] \times [\alpha + \{(1-\alpha) \times \frac{n}{n}\}] \times N^a_{j_1}, \hspace{2pt} 0 \hspace{2pt} \leq \hspace{2pt} j_1 \hspace{2pt} < \hspace{2pt} n 
\end{equation}

Note: for the case $j_1$ = $n$, $N^{Tr}_{n,n}$ = $N^a_n$ and we are done.

Now, equation \ref{proofata} can be written as:
\begin{equation}
	\label{proofata2}
	\begin{aligned}
		N^{Tr}_{j_1,n} &= \prod_{j = j_1}^{n - 1} \left[(1 - \alpha) - \left((1-\alpha) \times \frac{j}{n}\right)\right] \times N^a_{j_1}, & 0 &\leq j_1 < n \\
		&= \prod_{j = j_1}^{n - 2} \left[(1 - \alpha) - \left((1-\alpha) \times \frac{j}{n}\right)\right] \\
		&\times \left[(1 - \alpha) - \left((1-\alpha) \times \frac{n - 1}{n}\right)\right] \times N^a_{j_1}, & 0 &\leq j_1 < n \\
	\end{aligned}
\end{equation}

Then, we have:
\begin{equation}
\label{proofata3}
\begin{aligned}
    N^{Tr}_{j_1,(n - 1)} = &\prod_{j = j_1}^{n - 2} [(1 - \alpha) - \{(1-\alpha) \times \frac{j}{n}\}] \\
    &\times [\alpha + \{(1-\alpha) \times \frac{n - 1}{n}\}] \times N^a_{j_1}, \hspace{2pt} 0 \hspace{2pt} \leq \hspace{2pt} j_1 \hspace{2pt} < \hspace{2pt} (n - 1)
  \end{aligned}
\end{equation}
Adding equations \ref{proofata2} and \ref{proofata3}, we get:
\begin{equation}
\label{proofata4}
    \begin{aligned}
       N^{Tr}_{j_1,n} + N^{Tr}_{j_1,(n - 1)} &= \prod_{j = j_1}^{n - 2} [(1 - \alpha) - \{(1-\alpha) \times \frac{j}{n}\}] \\
        &\times N^a_{j_1}, \hspace{2pt} 0 \hspace{2pt} \leq \hspace{2pt} j_1 \hspace{2pt} < \hspace{2pt} (n - 1) \\
    &= \prod_{j = j_1}^{n - 3} [(1 - \alpha) - \{(1-\alpha) \times \frac{j}{n}\}] \\
    &\times [(1 - \alpha) - \{(1-\alpha) \times \frac{n - 2}{n}\}] \times N^a_{j_1}, \hspace{2pt} 0 \hspace{2pt} \leq \hspace{2pt} j_1 \hspace{2pt} < \hspace{2pt} (n - 1) \\
    \end{aligned}
\end{equation}
Also:
\begin{equation}
\label{proofata5}
\begin{aligned}
    N^{Tr}_{j_1,(n - 2)} = &\prod_{j = j_1}^{n - 3} [(1 - \alpha) - \{(1-\alpha) \times \frac{j}{n}\}] \\
    &\times [\alpha + \{(1-\alpha) \times \frac{n - 2}{n}\}] \times N^a_{j_1}, \hspace{2pt} 0 \hspace{2pt} \leq \hspace{2pt} j_1 \hspace{2pt} < \hspace{2pt} (n - 2) \\
    \end{aligned}
\end{equation}
Adding equations \ref{proofata4} and \ref{proofata5}, we get:
\begin{equation}
	\begin{aligned}
       N^{Tr}_{j_1,n} + N^{Tr}_{j_1,(n - 1)} + N^{Tr}_{j_1,(n - 2)} = &\prod_{j = j_1}^{n - 3} [(1 - \alpha) - \{(1-\alpha) \times \frac{j}{n}\}] \\
       &\times N^a_{j_1}, \hspace{2pt} 0 \hspace{2pt} \leq \hspace{2pt} j_1 \hspace{2pt} < \hspace{2pt} (n - 2) \\
       \end{aligned}
\end{equation}
Proceeding similarly as above, we get:
\begin{equation}
    \label{proofata6}
    \sum_{j=j_1 + 1}^{n} N^{Tr}_j = [(1 - \alpha) - \{(1-\alpha) \times \frac{j_1}{n}\}] \times N^a_{j_1}
\end{equation}
Adding equations \ref{proofata1} and \ref{proofata6}, we get:
\begin{equation*}
    \sum_{j=j_1}^{n} N^{Tr}_j = N^a_{j_1}
\end{equation*}
 
\section{Model Calibration, Verification and Validation}
\label{appvalid}

The verification and validation exercise for this model involves examining each of its modules for internal and external validity. We describe this process in this section. 

First, we consider the agent-based transmission dynamics component of the model. Validating this component would ideally involve comparing the key disease transmission outcomes across time generated by the model - the extent of the spread of the disease - against external longitudinal data. The relevant disease transmission outcomes in our case are: (a) the HCV antibody prevalence, which is the proportion of the population (agent cohort) who have ever been infected with HCV, including active and previous (cured) infections; (b) the HCV RNA prevalence, the proportion of the population with active HCV infections; and (c) the IDU prevalence, the proportion of the population actively engaging in injecting drug use. In the Indian context, to the best of our knowledge, longitudinal data for the spread of HCV in Punjab is not available; however, reliable cross-sectional data is available. Therefore, for validating this component, we compare the HCV antibody, RNA and the IDU prevalence estimates generated by the model at a suitably chosen time point with corresponding estimates from the relevant epidemiological literature. We chose the study by \cite{sood2018burden} for estimates of these outcomes, as it is possibly the largest cross-sectional study of the burden of HCV conducted for the state of Punjab. The study reported an HCV antibody prevalence of 3.6\% (95\% CI: 3.0 - 4.2\%), an HCV RNA prevalence of 2.6\% (2.0 - 3.1\%) and an IDU prevalence of 0.1\% (95\% CI not reported). 

At this stage, we recall that estimates of all input parameters for the transmission module were not available, and three parameters were designated as `calibration variables' to be estimated via the model calibration process. The calibration process involved running the simulation model for a period of 50 years and varying the values of these parameters systematically until the key transmission-related or epidemiological outcomes (the HCV antibody, RNA and IDU prevalences) of the model were suitably close to those mentioned above (henceforth referred to as calibration targets). Thus the validation process for these epidemiological model outcomes becomes the `calibration' process. 

There is more than one way to conduct the calibration process. For example, in a previous version of this model, the authors cast the calibration process for estimating the calibration variables as a discrete simulation optimization problem \citep{das2021discrete}. Guided by the calibration variable estimates obtained by \cite{das2021discrete}, we obtained estimates of the calibration variables via a manual search. Our stopping criterion was as follows. We conduct 30 replications of the simulation until the end of the calibration period (50 years) for a given set of values of the calibration variables. At the end of the calibration period, we calculate the average value of each epidemiological outcome, denoted by: $\hat{o}_1$ = average HCV antibody prevalence, $\hat{o}_2$ = average HCV RNA prevalence, and $\hat{o}_3$ = average IDU prevalence (here $\hat{o}_i = \frac{1}{n} \sum_{j = 1}^{30} \hat{o}_{ij}$). We then define the absolute value of the relative fractional deviation of each $\hat{o}_i ~(i \in\{1,2,3\})$ from its calibration target $o_i ~(i \in \{1,2,3\})$ as $d_i = \frac{|o_i - \hat{o}_i|}{o_i}, ~i \in \{1,2,3\}$. If the sum of the $d_i ~(i \in \{1,2,3\})$ is less than 0.2, indicating an average absolute fractional error of less than 7\% across the epidemiological outcomes, we terminate the calibration process. 

From the calibration process, the probability of acquiring an HCV infection from a contaminated medical environment, the probability of a non-IDU being converted to an IDU, and probability of IDUs engaging in needle-sharing during a group injecting event were estimated as $0.004$, $2.8 \times 10^{-5}$ and $0.41$, respectively. From the estimate of probability of acquiring an HCV infection from a contaminated medical environment, the probability of HCV transmission via a contaminated blood transfusion was estimated as 0.806 using equation~\ref{eqpbt} in section \ref{medtrans}. Based on the clinical literature \citep{chakravarti2013study}, it is evident that not every contaminated blood transfusion results in HCV transmission, and hence this estimate appears reasonable.

Model estimates of the key epidemiological outcomes at the end of the calibration period are provided in Table \ref{tab:calib}. We observe coefficients of variation of approximately 20\%, indicating relatively stable model performance. It is evident from the table that there is no statistically significant difference (at a 5\% level of significance) between the model estimates of the epidemiological outcomes and the calibration targets.
 
\begin{table}[htbp]
  \centering
  \caption{Key epidemiological outcomes obtained from the model calibration process. SD = standard deviation.}
\resizebox{\textwidth}{!}{
    \begin{tabular}{|c|c|c|}
    \hline
    \textbf{Epidemiological outcome} & \textbf{Calibration target (95\% CI)} & \textbf{Model outcome: mean (SD)} \\
    \hline
    HCV antibody prevalence & 3.60\% (3.0 - 4.2\%) & 3.57\% (0.75\%) \\
    \hline
    HCV RNA prevalence & 2.60\% (2.0 - 3.1\%) & 2.77\% (0.58\%) \\
    \hline
    IDU prevalence & 0.10\% (not reported) & 0.10\% (0.01\%) \\
    \hline
    \end{tabular}%
}
  \label{tab:calib}%
\end{table}%

An additional level of validation for the disease transmission component of the agent-based model, distinct from the calibration process, was carried out by comparing estimates of the contribution of the medical and social interaction environments to corresponding real-world observations in the Indian context reported by \cite{chakravarti2013study}. The authors reported on the provenance of HCV infections among a cohort in a North Indian state located close to Punjab. The authors reported that 74.1\% of the HCV infections they studied were caused due to unsafe medical procedures, 16.7\% due to injecting drug use and 9\% with tattooing. Given that infections due to tattoing were subsumed within the social interaction environment in our model, we compared the model estimates of the contributions of each environment to the total number of HCV infections across the calibration period to the corresponding contributions from the study by \cite{chakravarti2013study}. The proportion of HCV infections caused in the medical environment was estimated (from 30 replications) as 73.6\% with a standard deviation of 2.9\%. A one-sample $t$-test for equality of means did not find this difference (compared to 74.1\%) to be statistically significant at a significance level of 0.05. The contribution of the social interaction environment was estimated as 26.4\% (standard deviation of 2.9\%), against the value of 25.7\% (sum of the contributions of injecting drug use and tattooing) reported in \cite{chakravarti2013study}. Once again, this difference was not found to be statistically significant via a similar $t$-test.

The disease progression and non-DRD components of the simulation were validated by comparing outcomes from our models to outcomes from two studies conducted in the Indian context - \cite{aggarwal2017cost} and \cite{chugh2019real} - after using initial conditions similar to those used in these studies. 

\cite{aggarwal2017cost} evaluated the cost-effectiveness of HCV treatment with DAAs using a DTMC for modeling disease progression among a cohort of patients aged 35 years (transmission dynamics were not considered). They estimated the average life years lived by chronic non-cirrhotic patients (states $F0-F3$) to be 30.25 years, bringing the average life expectancy of these patients to 65.25 years. We used the HCV stage-wise initial distribution of patients from \cite{aggarwal2017cost}; however, we used the age-wise population distribution of Punjab in the year 2011 \citep{mha11}, when the last census was carried out, yielding a median agent age of 29 years. Under these conditions, our model estimated the average life years lived by patients from states $F0-F3$ to be 35.3 years, yielding an average life expectancy of 64.3 years for patients in HCV stages $F0-F3$. This is comparable to the estimate obtained by \cite{aggarwal2017cost}.

We also considered the study of \cite{chugh2019real} who perform a similar cost-effectiveness analysis with a different set of DAAs. The authors state that the average life expectancy of non-cirrhotic chronic patients is similar to that of the average life expectancy in the general population. Using the age-wise population distribution of \cite{chugh2019real} in our model, we estimated the average life years lived by non-cirrhotic chronic patients to be 38.3 years. This yielded an average life expectancy of 67.3 years, which corresponded to the reported life expectancy of India of 68.5 years in 2015 \citep{wpp2019census}.

Several other internal and external verification and validation checks were performed, in line with recommendations by \cite{sargent2010verification} for simulation models. These included extreme condition tests for key input parameters, validation of the demographic module, and validation of key model inputs and outputs by our expert clinical collaborator. 

\section{Outcomes Estimation: Additional Details}
\label{appcostqol}
The unit costs and resource use associated with HCV management are provided in Table~\ref{tabcost}. The utility weights capturing the age-related deterioration in quality of life by age group and HCV disease state are provided in Table~\ref{tabqol}.

\begin{table}[htbp]
	\centering
	\caption{Utility weights or quality-of life multipliers according to disease state and age group. The disease state utility weights were obtained from \cite{chugh2019real} and the age-based utility weights from \cite{he2016prevention}.}
	\resizebox{\textwidth}{!}{
		\begin{tabular}{|c|c|c|c|c|c|}
			\hline
			& \textbf{HCV disease state} & \textbf{Utility weight} &       & \textbf{Age group (in years)} & \textbf{Utility weight} \\
			\cline{2-3}\cline{5-6}          & F0-F3 & 0.63  &       & 0-29  & 0.921 \\
			\cline{2-3}\cline{5-6}          & F4    & 0.56  &       & 30-39 & 0.906 \\
			\cline{2-3}\cline{5-6}    Disease & DC    & 0.44  & Age-based & 40-49 & 0.875 \\
			\cline{2-3}\cline{5-6}    state & HCC   & 0.44  & utility weight & 50-59 & 0.849 \\
			\cline{2-3}\cline{5-6}    utility weight & LT    & 0.84  &       & 60-69 & 0.826 \\
			\cline{2-3}\cline{5-6}          & SVR2  & 0.93  &       & 70-79 & 0.786 \\
			\cline{2-3}\cline{5-6}          & Healthy & 1     &       & 80+   & 0.753 \\
			\hline
		\end{tabular}%
	}
	\label{tabqol}%
\end{table}%

%\newpage
\begin{table}[htbp]
	\centering
	\caption{Unit costs and resource use for HCV management. All costs are in INR.}
	\footnotesize
	\begin{threeparttable}
		\begin{tabular}{|C{2.5cm}|c|c|c|C{2cm}|}
			\hline
			\textbf{F0-F3} & \textbf{F4} & \textbf{DC} & \textbf{HCC} & \textbf{LT} \\
			\hline
			\multicolumn{5}{|c|}{\textbf{One-time cost heads and values}} \\
			\hline
			CBC, 200 & CBC, 200 & CBC, 200 & Surgery, 150000 & Transplant, 2000000 \\
			LFT, 500 & RBS, 100 & RBS, 100 & RFA (30\% of patients), 150000 & Others, 500000 \\
			RBS, 100 & KFT, 300 & TSH, 200 & TACE (30\% of patients), 150000 &  \\
			KFT, 300 & TSH, 200 & HbsAg, 300 & MWA (30\% of patients), 150000 &  \\
			TSH, 200 & HbsAg, 300 & Anti-HIV, 300 & Radiation (5\% of patients), 100000 &  \\
			HbsAg, 300 & Anti-HIV, 300 & Fibroscan, 1500 & TARE (1\% of patients), 750000 &  \\
			Anti-HIV, 300 & Fibroscan, 1500 &       & Costs from heads in DC state, 2600 &  \\
			USG 500 &       &       &       &  \\
			Fibroscan, 1500 &       &       &       &  \\
			\hline
			Net: 3900 & Net: 2900 & Net: 2600 & Net: 300100 & Net: 2500000 \\
			\hline
			\multicolumn{5}{|c|}{\textbf{Recurring cost heads, annual frequency and values}} \\
			\hline
			Doctor visits, 6, 500 & Doctor visits, 6, 500 & Doctor visits, 6, 500 & Drugs (30\% of patients), 50000 & Drugs and others, 24000 \\
			PPIs, 8 courses, 1000 & PPIs, 8 courses, 1000 & PPIs, 8 courses, 1000 & BSC (40\% of patients), 12, 2500 &  \\
			& USG, 2, 500 & USG, 2, 500 & Costs from heads in DC state, 23600 &  \\
			& LFT, 2, 500 & LFT, 2, 500 &       &  \\
			& Hospitalization (10\% & KFT, 4, 100 &       &  \\
			& of patients), 2, 25000 & Hospitalization (10\% &       &  \\
			& Endoscopy, 1, 3000 & of patients), 2, 25000 &       &  \\
			& AFP, 1, 700 & Endoscopy, 1, 3000 &       &  \\
			&       & PTI, 4, 100 &       &  \\
			\hline
			Net: 11000 & Net: 21700 & Net: 23600 & Net: 50600 & Net: 24000 \\
			\hline
		\end{tabular}%
		\begin{tablenotes}
			\item[] \textit{Notes.} INR: Indian Rupee, CBC: complete blood count, LFT: liver function test, RBS: random blood sugar, KFT: kidney function test, TSH: thyroid stimulating hormone, HbsAg: surface antigen of hepatitis B virus (HBV), anti-HIV: HIV antibody test, USG: ultrasonography, RFA: radiofrequency ablation, TACE: transarterial chemoembolization, MWA: microwave ablation, TARE: background transarterial chemoembolization, PPIs: proton pump inhibitors, AFP: alpha-fetoprotein, PTI: prothrombin-time, BSC: best supportive care.
		\end{tablenotes}
	\end{threeparttable}
	\label{tabcost}%
\end{table}%
\normalsize

 \end{document}